\documentclass[letterpaper,twocolumn,american,prx,amsmath,amssymb,superscriptaddress,nofootinbib]{revtex4-2}
\usepackage[T1]{fontenc}
\usepackage{babel}
\usepackage{prettyref}
\usepackage{dsfont}
\usepackage{amsmath}
\usepackage{graphicx}
\usepackage{float}
\usepackage[unicode=true,
 bookmarks=true,bookmarksnumbered=false,bookmarksopen=false,
 breaklinks=false,pdfborder={0 0 1},backref=false,colorlinks=false]
 {hyperref}
\hypersetup{pdftitle={Learning actions from Data using Invertible Neural Networks},
 pdfauthor={Claudia Merger, Alexandre Rene, Kirsten Fischer, Peter Bouss, Sandra Nestler, David Dahmen, Carsten Honerkamp, Moritz Helias},
 colorlinks=true,linkcolor=black,citecolor=black,urlcolor=black,filecolor=black}

\makeatletter

\pdfpageheight\paperheight
\pdfpagewidth\paperwidth

\providecommand{\tabularnewline}{\\}

\usepackage{tikz}
\usetikzlibrary{calc}

\usepackage[latin1]{inputenc}
\usepackage{amsmath}
\usepackage{prettyref}
\newrefformat{eq}{\hyperref[#1]{Eq.~(\ref{#1})}}
\newrefformat{cap}{\hyperref[#1]{Fig.~\ref{#1}}}
\newrefformat{fig}{\hyperref[#1]{Fig.~\ref{#1}}}
\newrefformat{tab}{\hyperref[#1]{Table ~\ref{#1}}}
\newrefformat{sec}{\hyperref[#1]{Section~\ref{#1}}}
\newrefformat{subsec}{\hyperref[#1]{Section~\ref{#1}}}
\newrefformat{cha}{\hyperref[#1]{Chapter~\ref{#1}}}
\newrefformat{app}{\hyperref[#1]{Appendix~\ref{#1}}}

\usepackage{MnSymbol}
\DeclareMathAlphabet\mathbb{U}{msb}{m}{n}

\usepackage{booktabs}
\usepackage[title]{appendix}

\usepackage[force]{feynmp-auto}

\makeatother

\begin{document}
\global\long\def\R{\mathbb{R}}%
\global\long\def\llangle{\langle\!\langle}%
\global\long\def\rrangle{\rangle\!\rangle}%
\global\long\def\T{\mathrm{\mathrm{T}}}%
\global\long\def\tran{\mathrm{^{\mathrm{T}}}}%
\global\long\def\lowerindex{\vphantom{X^{N}}}%
\global\long\def\extralowerindex{\vphantom{X^{N^{N}}}}%
\global\long\def\symm{\operatorname{sym}}%
\global\long\def\DPath{\operatorname{Dpath}}%
\global\long\def\Tr{\operatorname{Tr}}%

\title{Spurious self-feedback of mean-field predictions inflates infection
curves}
\author{Claudia Merger}
\email{c.merger@fz-juelich.de}

\affiliation{Institute of Neuroscience and Medicine (INM-6) and Institute for Advanced
Simulation (IAS-6) and JARA-Institute Brain Structure-Function Relationships
(INM-10), Jülich Research Centre, Jülich, Germany}
\affiliation{RWTH Aachen University, Aachen, Germany}
\author{Jasper Albers}
\affiliation{Institute of Neuroscience and Medicine (INM-6) and Institute for Advanced
Simulation (IAS-6) and JARA-Institute Brain Structure-Function Relationships
(INM-10), Jülich Research Centre, Jülich, Germany}
\affiliation{RWTH Aachen University, Aachen, Germany}
\author{Carsten Honerkamp}
\affiliation{RWTH Aachen University, Aachen, Germany}
\author{Moritz Helias}
\affiliation{Institute of Neuroscience and Medicine (INM-6) and Institute for Advanced
Simulation (IAS-6) and JARA-Institute Brain Structure-Function Relationships
(INM-10), Jülich Research Centre, Jülich, Germany}
\affiliation{RWTH Aachen University, Aachen, Germany}
\date{\today}
\begin{abstract}
The susceptible-infected-recovered (SIR) model and its variants form
the foundation of our understanding of the spread of diseases. Here,
each agent can be in one of three states (susceptible, infected, or
recovered), and transitions between these states follow a stochastic
process. The probability of an agent becoming infected depends on
the number of its infected neighbors, hence all agents are correlated.
A common mean-field theory of the same stochastic process however,
assumes that the agents are statistically independent. This leads
to a self-feedback effect in the approximation: when an agent infects
its neighbors, this infection may subsequently travel back to the
original agent at a later time, leading to a self-infection of the
agent which is not present in the underlying stochastic process. We
here compute the first order correction to the mean-field assumption,
which takes fluctuations up to second order in the interaction strength
into account. We find that it cancels the self-feedback effect, leading
to smaller infection rates. In the SIR model and in the SIRS model,
the correction significantly improves predictions. In particular,
it captures how sparsity dampens the spread of the disease: this indicates
that reducing the number of contacts is more effective than predicted
by mean-field models.
\end{abstract}
\maketitle

\section{Introduction\label{sec:introduction}}

The Susceptible-Infected-Recovered (SIR) models and its variants
\citep{kermackContributionsMathematicalTheory1991} are highly popular
models for the spread of diseases. They are built on the assumption
that, to model the spread of a disease in a large population, the
effect of each single individual is small. The stochastic SIR model
instead describes the relative fractions of individuals that are in
either of the three states, susceptible, infected, or recovered. Changes
of these fractions are controlled by the average probability $\beta$
of infecting a contact person and the average probability of recovery
$\mu$. One then assumes the same average transition rates between
states for each individual, which only depend on the size of the relative
fractions of the population in each states. Assuming homogeneous interactions
between all individuals, and averaging over the stochastic process
one arrives at a set of three coupled differential equations, which
describe the average growth of the infection curve.

However, it has been demonstrated that the assumption of homogeneity
of interactions across individuals and time is insufficient \citep{thurner_network-based_2020,heltberg_spatial_2022,tkachenko_time-dependent_2021,newman_spread_2002,moreno_epidemic_2002,may_infection_2001},
and that real world propagation must be modeled using heterogeneous
networks. One then describes a stochastic process of $N$ agents,
whose probability of infection depends on the number of infected nearest
neighbors on a contact graph. This is also referred to as individual-based
modeling (IBM) or agent-based modeling (ABM) \citep{pastor-satorrasEpidemicProcessesComplex2015b}.
Due to the interaction of all agents, the average growth of the infection
curve cannot be computed. Under the assumption that the agents statistics
are independent, one arrives at a set of $3N$ coupled update equations,
the mean-field approximation of the stochastic process that describe
the average infection curve for each agent.

The mean-field equations are an insufficient approximation of the
stochastic process, especially if the contact graph is sparse \citep{thurner_network-based_2020}.
We will demonstrate this here by comparison to simulations of the
same process, finding that mean-field predictions invariably overestimate
the number of infected agents. The mismatch between the ground truth
and its mean-field approximation is most prominent in sparse network
structures. We then show that the predictions can be improved in a
systematic manner using a dynamical Plefka expansion analogous to
\citep{roudi_dynamical_2011}. We treat both the SIR and SIRS model.
The latter is a variant of the former. Here, agents may lose their
immunity after recovery, so that multiple infections of the same agent
are possible. In the case of the SIR model, the corrections obtained
via this expansion  cancel a spurious self-feedback effect: when
an agent infects its neighbors, this infection may travel back to
the original agent, leading to a self-infection of the agent at a
later time. This self-feedback effect is present in the mean-field
approximation of the model, but not in the underlying stochastic
process. We show that the corrected equations can predict the dampening
effect of sparsity both in the SIR and SIRS models. Additionally,
computing the corrected theory is efficient, requiring one additional
dynamical variable per node, such that the number of dynamical variables
which one has to track still only scales with $N$.

Apart from the mean-field method, there exist further methods to approximate
the dynamics. They are dynamical message-passing (DMP) \citep{karrer_message_2010,lokhov_inferring_2014}
and what is known as the pair approximation \citep{mata_pair_2013}.
Both of these approaches yield a higher-dimensional set of coupled
differential equations than mean-field. For DMP, one must track one
dynamical variable per edge on the contact graph, for the pair approximation
one must track one dynamical variable per pair of nodes, thus $\mathcal{\mathcal{O}}\left(N^{2}\right)$
variables. We further discuss the relation of the fluctuation correction
to DMP and the pair approximation in \prettyref{sec:Discussion}.

This paper is organized as follows: in \prettyref{sec:SIR-model},
we introduce the SIR model and the most common approximation of the
underlying stochastic process. In \prettyref{sec:Cancellation-of-self-feedback},
we present the fluctuation correction to the SIR model and its variants.
We demonstrate the effect of the self-feedback term in \prettyref{sec:Experiments}.
Finally we discuss our result in \prettyref{sec:Discussion}.

\section{SIR model\label{sec:SIR-model}}

We model $N$ individuals $i$ whose social contacts are represented
by a graph with adjacency matrix $a$. Individuals may only transmit
the disease to or be infected by their direct nearest neighbors on
the graph. We assign to each individual a set of two binary variables,
$S_{i},I_{i}\in\left\{ 0,1\right\} $. The susceptible state corresponds
to $S_{i}=1$ and $I_{i}=0$, the infected state is defined by $S_{i}=0$
and $I_{i}=1$. Once an individual recovers, we set $S_{i}=I_{i}=0$.
All other configurations of $S_{i},I_{i}$ are not allowed. At each
time step, the probability of infection of node $i$ is $\phi(\theta_{i})$
with the input $\theta_{i}$ to node $i$ defined by
\begin{equation}
\theta_{i}(t)=h_{i}(t)+\beta\sum_{j}a_{ij}I_{j}(t)\label{eq:input_field_theta}
\end{equation}
with an external field $h_{i}(t)$ (corresponding to influx of infections
from outside the network) and the interaction strength $\beta$, which
increases the probability of an infected node infecting a susceptible
one. We will later set $h_{i}=0$ to investigate the endogenous dynamics.
In principle, $\phi:[0,\infty)\rightarrow[0,1]$ can be any function
which maps to a probability. We will further assume that, close to
the origin, $\phi$ is well approximated by the identity, such that
$\phi(0)=0$, and $\phi^{\prime}(0)=1$. The standard formulation
of the SIR model defines $\phi(\theta)=\theta,$ which we will adopt
here and in the following. In this case, the interaction strength
$\beta$ is just the probability of an infected node transmitting
the disease to one of its nearest neighbors. An infected individual
recovers in each time step with probability $\mu$.

\begin{figure}
\begin{centering}
\includegraphics[width=8.6cm]{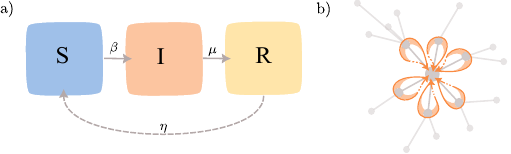}
\par\end{centering}
\caption{\label{fig:Transition-probabilities} a) Transition probabilities
in the SIR (solid lines) and SIRS (solid and dashed lines) model.
b) Illustration of self-feedback.}
\end{figure}

In the SIRS model, a variant of the SIR model with waning immunity,
recovered individuals may move back to the susceptible state with
probability $\eta$. An overview of the transition probabilities between
the compartments is shown in \prettyref{fig:Transition-probabilities}.

The model is a stochastic Markov process because transitions take
place randomly with the transition probabilities defined above and,
given the current state, the evolution is history independent. We
denote as $\langle\ldots\rangle$ the expectation value over this
randomness. Because the state variables $S_{i}$ and $I_{i}$ are
binary $\in\{0,1\}$, their expectation values $\langle S_{i}\rangle=:\rho_{i}^{S}$
and $\langle I_{i}\rangle=:\rho_{i}^{I}$ equal the probabilities
to occupy the corresponding state, $S$ or $I$, respectively. We
are interested in the evolution of these probabilities $\rho_{i}^{\alpha}$
that individual $i$ is in state $\alpha\in\left\{ S,I,R\right\} $.
Taking all possible transitions together, we may write down the difference
$\Delta\rho_{i}^{\alpha}(t+1)=\rho_{i}^{\alpha}(t+1)-\rho_{i}^{\alpha}(t)$
between to time steps exactly:

\begin{align}
\Delta\rho_{i}^{S}(t+1) & =\eta\rho_{i}^{R}(t)-\left\langle S_{i}(t)\theta_{i}(t)\right\rangle \nonumber \\
\Delta\rho_{i}^{I}(t+1) & =-\mu\rho_{i}^{I}(t)+\left\langle S_{i}(t)\theta_{i}(t)\right\rangle \,.\label{eq:SIR_dynamics_exact}
\end{align}
These update equations can be seen as a balance of probability influx
and outflux from the states: individuals enter the susceptible from
the recovered state with probability $\eta$, which yields the first
term in $\Delta\rho_{i}^{S}(t+1)$. On the other hand, individual
agents leave the state if they become infected which happens with
probability $\left\langle S_{i}(t)\theta_{i}(t)\right\rangle $, which
may be interpreted as the product of the probability $\theta_{i}$
to become infected given the individual is susceptible times the probability
to be susceptible. This gives us the second term in $\Delta\rho_{i}^{S}(t+1)$
and the second term in $\Delta\rho_{i}^{I}(t+1)$. Infected agents
leave the infected state with the probability $\mu$ for recovery,
which accounts for the first term in $\Delta\rho_{i}^{I}(t+1)$.
Here and in the following, setting $\eta=0$ yields the evolution
for the SIR model.It is sufficient to track only $\rho_{i}^{I}$ and
$\rho_{i}^{S}$ since $\rho_{i}^{R}$ follows from $\rho_{i}^{R}=1-\rho_{i}^{S}-\rho_{i}^{I}$.

The difficulty in evaluating \prettyref{eq:SIR_dynamics_exact} lies
in computing the expectation value $\left\langle S_{i}(t)\theta_{i}(t)\right\rangle $,
for which we must compute the cross-correlations between individuals
for all times. As the system size grows, an evaluation of the exact
update equations for $\rho_{i}^{\alpha}$ becomes computationally
infeasible. A typical approximation stipulates that individuals are
statistically independent, meaning that averages of the type $\left\langle S_{i}(t)\theta_{i}(t)\right\rangle $
factorize into $\left\langle S_{i}(t)\right\rangle \left\langle \theta_{i}(t)\right\rangle $,
since $\theta_{i}$ neither depends on $S_{i}$ nor $I_{i}$. This
yields the individual-based mean-field equations:
\begin{align}
\Delta\rho_{i}^{S}(t+1) & =\eta\rho_{i}^{R}(t)-\rho_{i}^{S}(t)\beta\sum_{j}a_{ij}\rho_{j}^{I}(t)\nonumber \\
\Delta\rho_{i}^{I}(t+1) & =-\mu\rho_{i}^{I}(t)+\rho_{i}^{S}(t)\beta\sum_{j}a_{ij}\rho_{j}^{I}(t)\,,\label{eq:SIR_dynamics_MF}
\end{align}
which, apart from replacing $\left\langle S_{i}(t)\theta_{i}(t)\right\rangle $
by $\rho_{i}^{S}(t)\beta\sum_{j}a_{ij}\rho_{j}^{I}(t),$ are equal
to \prettyref{eq:SIR_dynamics_exact}. We will refer to them simply
as mean-field equations in the following. Given initial expectation
values $\rho_{i}^{\alpha}(0)$, these equations can then be iterated
to compute the prediction for the average infection curve. However,
we will show in the following that the assumption of statistical independence
is not well-justified. Note that in \prettyref{eq:SIR_dynamics_MF},
the probability of a nearest neighbor of node $i$ being infected,
$\rho_{j}^{I}(t)$ contains the term $\rho_{j}^{S}(t-1)\beta a_{ji}\rho_{i}^{I}(t-1)$,
which then contributes to the probability of node $i$ being infected
two time steps later. In the next section, we will see that this term
is canceled by fluctuations.

\section{Cancellation of self-feedback\label{sec:Cancellation-of-self-feedback}}

Indeed, we find that correlations do play a role. We show in \prettyref{app:Expansion_of_effective_action}
that the mean-field Eqs. \eqref{eq:SIR_dynamics_MF} can be obtained
as a first order approximation in $\beta$ from a systematic fluctuation
expansion: we extend the individual-based mean-field equations by
performing a Plefka expansion \citep{Plefka82_1971} to second order.
In \citep{roudi_dynamical_2011}, Roudi and Hertz demonstrated how
to do the Plefka expansion in the dynamical setting for a system of
Ising spins. The resulting update equations are also known as dynamical
TAP equations, a non-equilibrium version of the so-called TAP correction
term \citep{Thouless77_593}, first derived in \citep{Vasiliev74}.
Our calculation closely follows \citep{roudi_dynamical_2011}, with
the important distinction that here the transition probability between
states depends on the state itself. The terms in the expansion are
ordered by powers of $\text{\ensuremath{\beta}}$, hence we expand
around the point where $\beta=0$, in which case the independence
of individuals is exact, and the model can be solved exactly. To first
order, one obtains Eqs. \eqref{eq:SIR_dynamics_MF}, and to second
order, each update equation acquires a correction term of order $\beta^{2},$
which reads 
\begin{align}
\Delta\rho_{i}^{S}(t+1) & =\eta\rho_{i}^{R}(t)-\rho_{i}^{S}(t)\beta\sum_{j}a_{ij}\left(\rho_{j}^{I}(t)-\rho_{j\text{\ensuremath{\leftarrow i}}}^{I}(t)\right),\nonumber \\
\Delta\rho_{i}^{I}(t+1) & =-\mu\rho_{i}^{I}(t)+\rho_{i}^{S}(t)\beta\sum_{j}a_{ij}\left(\rho_{j}^{I}(t)-\rho_{j\text{\ensuremath{\leftarrow i}}}^{I}(t)\right)\,,\label{eq:SIR_dynamics_TAP}
\end{align}
where $\rho_{j\text{\ensuremath{\leftarrow i}}}^{I}(t)$, to first
order in $\beta$, is just the probability that node $j$ was infected
by node $i$ at an earlier point in time and stayed in this state
since then,
\[
\rho_{j\text{\ensuremath{\leftarrow i}}}^{I}(t):=\beta a_{ji}\sum_{t^{\prime}\leq t-1}\rho_{j}^{S}(t^{\prime})\rho_{i}^{I}(t^{\prime})(1-\mu)^{t-t^{\prime}-1}\,.
\]
Here, $\rho_{i}^{I}(t^{\prime})\beta a_{ji}\rho_{j}^{S}(t^{\prime})$
describes the probability that node $j$ becomes infected at $t^{\prime}+1$
due to node $i$, and $(1-\mu)^{t-t^{\prime}-1}$ is the probability
that node $j$ does not recover between $t^{\prime}+1$ and $t$.
We find that the second order correction therefore is exactly the
term which cancels the self-feedback of $i$ onto itself to first
order in $\beta$. In the SIR model, this cancellation is necessary,
since in the stochastic process, any individual can either be susceptible
or infected, and nothing in between. Therefore, once an individual
has been infected, this infection may not travel back and re-infect
the individual, since the transitions $I\rightarrow S$ and $R\rightarrow S,R\rightarrow I$
have probability zero. Alternatively, one may view approximations
leading to Eqs. \eqref{eq:SIR_dynamics_MF} as artificially introducing
a positive self-feedback, which is not present in the underlying stochastic
process.

We performed the expansion to second order for the SIR, SIRS, and
the SIS model (see \prettyref{app:Expansion_of_effective_action}
and \prettyref{app:Extensions-to-SIS-and-SIRS}). In the latter, no
recovered state exists, rather infected agents move to the susceptible
state again with probability $\bar{\mu}$. Here and in the following,
we analyze only the SIR and SIRS model, an overview of the corrected
dynamical equations including the SIS model is found in \prettyref{app:overview}.

\section{Dampening effect of sparsity\label{sec:Experiments}}

We will now compare the predictions of Eqs. \eqref{eq:SIR_dynamics_MF}
to the corresponding results from Eqs. \eqref{eq:SIR_dynamics_TAP}
on different network topologies. We implement simulations of the SIR
and SIRS dynamics using NEST \citep{gewaltig_nest_2007}, a simulator
for spiking neural networks, which we adapt to encode transitions
between binary states, see \prettyref{app:NEST-implementation}.
The mean-field and TAP predictions were obtained by iterating the
update equations, see \prettyref{app:Mean-field-and-TAP}.

\subsubsection{SIR model}

\begin{figure*}
\centering{} \includegraphics[width=1\textwidth]{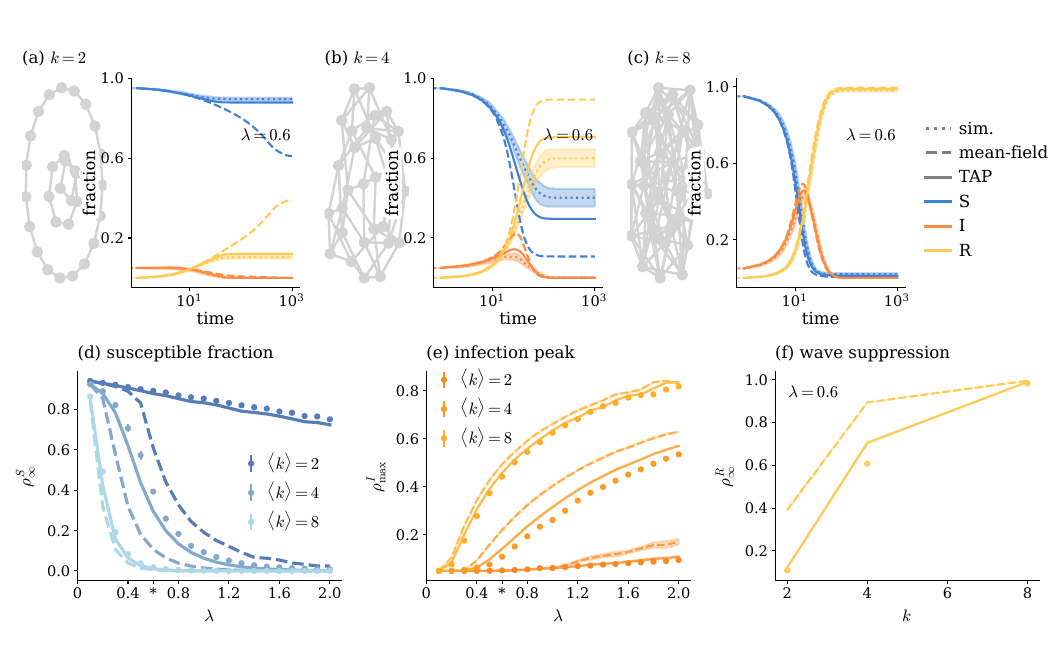}
\caption{\textbf{Sparsity suppresses spread of infection in the SIR model on
random regular networks.} \textbf{(a)}-\textbf{(c)} Dotted curves
show simulation results for fixed $\mu=10^{-1},$ and $\lambda:=\frac{\beta}{\mu}=0.6$,
shaded areas show one standard deviation, averaged over $10$ different
realizations of the stochastic infection process on the same network
realization with the same initially infected agents. Dashed curves
show mean-field result; solid curves the prediction from the TAP equations,
the second order correction. \textbf{(d)} Fraction of agents in the
susceptible state after the dynamics have stopped. Dots are simulation
results, curves show mean-field and TAP, shaded areas indicate one
standard deviation.  \textbf{(e)} Same as (d), but for the peak of
the infection curve. Stars along the axis mark the value of $\lambda$
used in (a)-(c). \textbf{(f)} Fraction of agents in the recovered
state after the dynamics have stopped over the degree $k$ of the
network. All results were obtained on networks with average size $N=10^{3}$.
All data in (d)-(f) are averaged over $10$ realizations of the adjacency
matrix, results in (a)-(c) stem from a single realization of the adjacency
matrix.}
\label{fig: Sparsity_supresses_spread}
\end{figure*}

We here study the effect which the average connectivity has on the
spread of the infection. To do so, we generate random regular graphs
of different degree $k$ per node, infect a small fraction ($5\%$)
of randomly chosen agents, and then let the dynamics evolve. Random
regular graphs possess no hierarchical structure: all nodes have equal
degree, and their connections are chosen at random. Hence, they are
suited to study the effect of the average connectivity on the spread
of the infection.

We compare the predictions of the mean-field and second-order approximation
to simulations of the stochastic process. In \prettyref{fig: Sparsity_supresses_spread}
(a)-(c) we show the infection curves, averaged over the whole population,
as well as a small example graph on the left hand side. (Note that
for $k=2$, graphs typically consist of periodic linear chains.) For
densely connected networks, the predictions of mean-field theory,
the second order correction and the simulation all agree. As the networks
become more sparse, however, the number of infections decreases much
faster than mean-field theory predicts, but is still well approximated
by the second order corrected theory. This shows that self-feedback
is highly relevant for sparse networks.

The latter observation holds true over a broad range of fractions
$\lambda=\frac{\beta}{\mu}$. After the infection wave has died down,
all agents must be either susceptible or recovered. As we report in
\prettyref{fig: Sparsity_supresses_spread} (d), the number of susceptible
agents in the final state, $\rho_{\infty}^{S}$, decreases with increasing
$\lambda$ -- but much slower than predicted by mean-field theory.
We find that, on sparse graphs in particular, the difference between
the mean-field prediction and the simulation is large, while \eqref{eq:SIR_dynamics_TAP}
captures the dynamics well.

Finally, we compare the height of the peak for different network topologies
and different values of $\lambda$ in \prettyref{fig: Sparsity_supresses_spread}
(e). We find that below a certain value of $\lambda$, which depends
on the degree $k$, the number of infected individuals does not grow
beyond the number of initially infected agents, thus the infection
wave dies out immediately. Beyond this point, the height of the infection
peak increases with $\lambda$. Again, the mean-field theory predicts
a higher value than observed in simulations. We find that the number
of infected agents overall decreases rapidly with the average degree.
After the dynamics have stopped, the number of recovered agents is
equivalent to the overall number of agents that experienced an infection.
In \prettyref{fig: Sparsity_supresses_spread}(f), we compare the
fraction $\rho_{\infty}^{R}$ of recovered agents after the infection
wave has died out to predictions from both theories, finding again
that as the graph becomes sparser, far fewer agents become infected
than predicted by mean-field theory.

Despite the good agreement between simulations and the second-order
correction \prettyref{eq:SIR_dynamics_TAP}, we find that the corrected
theory consistently overestimates the fraction of infected agents
slightly, both at the height of the wave, as shown in \prettyref{fig: Sparsity_supresses_spread}
(e), and in the average over the time trajectory see \prettyref{fig: Sparsity_supresses_spread}
(d). This is so, because the correction \prettyref{eq:SIR_dynamics_TAP}
still allows higher order feedback loops. Nevertheless, in comparison
to the difference between mean-field theory and the second order correction,
these higher-order terms appear to play a minor role.

In conclusion, the analysis confirms that sparser connectivity has
a more pronounced effect than mean-field theory suggests: The activity
dies out faster and its peak is much lower, hence the number of individuals
which have ever been infected is more strongly suppressed by a reduction
in the number of contacts.

\subsubsection{SIRS model}

In the SIRS model, an infected agent can reach the susceptible stage
again. Hence, in principle, a self-infection loop is allowed, provided
that the transition back to the susceptible state has taken place
in the meantime. Given this, the correction \prettyref{eq:SIR_dynamics_TAP}
seems counter-intuitive: clearly, self-feedback may exist in the SIRS
model, but the correction appears to cancel it. However, the allowed
self-feedback loop in the SIRS model must come with a factor of $\eta$,
which appears neither in the spurious self-feedback term in mean-field
theory, nor in the correction in \prettyref{eq:SIR_dynamics_TAP}.
The cancellation in \prettyref{eq:SIR_dynamics_TAP} is only partial;
higher order terms may indeed constitute self-feedback in the corrected
update equations.

\begin{figure*}
\centering{} \includegraphics[width=1\textwidth]{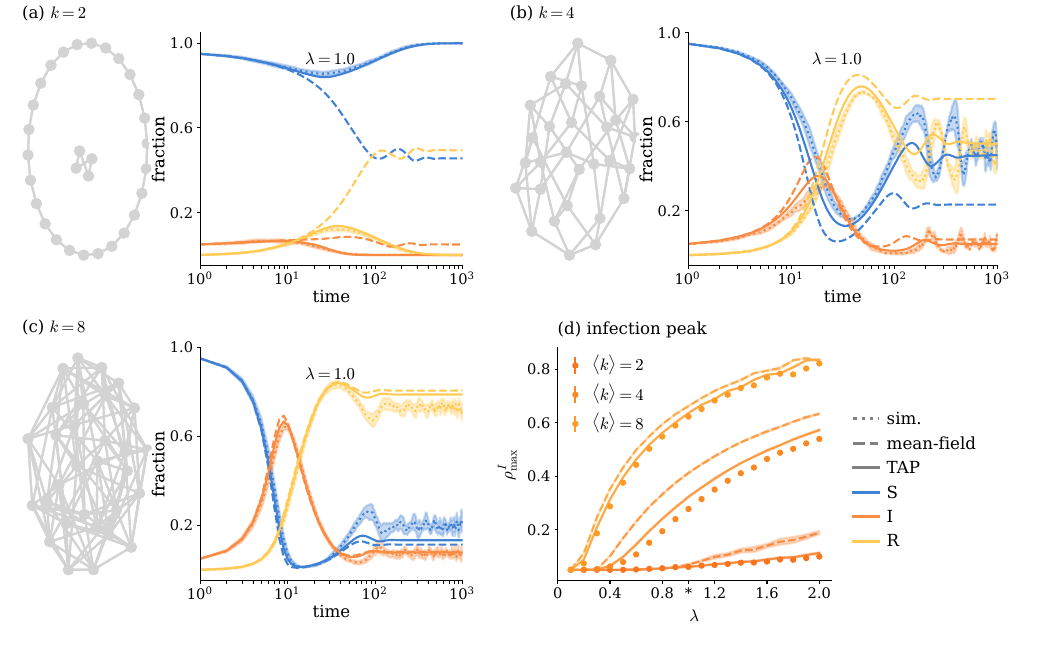}
\caption{\textbf{Fluctuation correction for SIRS model on random regular networks
with slowly waning immunity at system size $N=10^{3}$.} \textbf{(a)}-\textbf{(c)}
Dotted curves show results of simulations for fixed $\mu=10^{-1}$,
$\eta=10^{-2}$, and $\lambda=1.0$, shaded areas show one standard
deviation, averaged over the surviving runs of $10$ different realizations
of the stochastic process on the same network realization with the
same initially infected agents. Dashed curves show mean-field result;
solid curves the prediction from TAP equations, the second order correction.
\textbf{(d)} Peak of the infection curve, averaged over all runs (surviving-
and non-surviving). Dots show simulation result (error bars are typically
smaller than marker size). Dashed curves show the peak of the mean-field
prediction, solid curves the peak of the correction. All data in (d)
are averaged over $10$ realizations of the adjacency matrix. Stars
along the x-axis mark the value of $\lambda$ used in (a)-(c).}
\label{fig: SIRS_slowly_Waning}
\end{figure*}

In contrast to the SIR model, an endemic state, which is a state with
finite activity ($\rho^{I}>0$), can emerge in the SIRS model; then
the infection wave does not die out fully, rather, at all times, a
finite fraction of the population is infected. The fraction $\lambda_{c}$,
which marks the transition between the regimes where the activity
dies out or remains finite, is called the epidemic threshold. In \prettyref{app:Predictions-threshold},
we derive an estimator for this regime from \prettyref{eq:SIR_dynamics_TAP}
and apply it to scale-free networks, a class of strongly heterogeneous
networks, in \prettyref{app:Epidemic-threshold-scale-free}. Here,
we again focus on the effect of the average connectivity, and investigate
the spread of disease on random regular networks. We apply \prettyref{eq:SIR_dynamics_TAP}
to a system with slowly waning immunity; setting $\mu=10^{-1}$, $\eta=10^{-2}$
and varying $\beta$ between $\eta$ and $2\mu$. We show the results
of this procedure in \prettyref{fig: SIRS_slowly_Waning}.

We find that the fluctuation correction reproduces the average dynamics
well (see \prettyref{fig: SIRS_slowly_Waning} (a) - (c)). In particular,
in the sparse connectivity model, the infection wave dies out completely,
which is accurately predicted by the fluctuation correction, whereas
mean-field theory predicts an endemic state with large connectivity.
For large connectivity, the difference between mean-field and fluctuation
correction is also small; both overestimate the number of infections
slightly. The height of the infection peak is predicted with high
accuracy by the fluctuation correction at all levels of connectivity
(see~e \prettyref{fig: SIRS_slowly_Waning} (d)), while it is overestimated
by the mean-field result, especially for sparse graphs. For intermediate
and dense connectivity, here $k=4$ and $k=8$, we find that strong
fluctuations emerge in the dynamics, which are modeled neither by
mean-field, nor by the TAP equations. However, the fluctuation correction
overestimates the activity in the endemic state: it underestimates
the final fraction of susceptible individuals, while it overestimates
the number of infected agents in the final epidemic state, see \prettyref{fig: SIRS_slowly_Waning}
(c), (d).

\begin{figure*}
\begin{centering}
\includegraphics{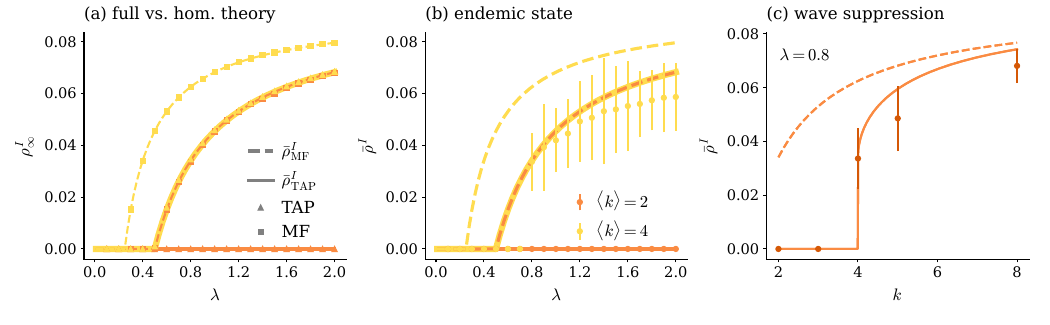}
\par\end{centering}
\caption{\textbf{Endemic state in the SIRS model on random regular networks
with} \textbf{slowly waning immunity, $\eta=10^{-2}$, and system
size $N=10^{3}$. (a) }Comparison of theoretical prediction for fraction
of infected agents at equilibrium for degree $k=2$ (orange) and $k=4$
(yellow). Triangles and squares show the outcome of the mean-field
theory (Eqs. \eqref{eq:SIR_dynamics_MF}) and TAP theory (Eqs. \eqref{eq:SIR_dynamics_TAP}),
respectively. Dashed and solid curves show theoretical results under
the assumption of homogeneity, Eqs. \eqref{eq:rho_bar_I_MF} and \eqref{eq:rho_bar_I_TAP},
respectively. Results from TAP theory for $k=4$ overlap with results
from mean-field theory for $k=2$. \textbf{(b)} Dashed and solid
curves show theoretical results under the assumption of homogeneity.
Dots are simulation results averaged over surviving runs and $10$
realizations of the adjacency matrix. Additionally, to mitigate the
effect of oscillations, results were averaged over the last $500$
steps of the simulation, comp. \prettyref{fig: SIRS_slowly_Waning}.
\textbf{(c)} Same as (b), but for fixed $\lambda=0.8$ and varying
degree $k$.}

\label{fig:SIRS_rhoI_endemic}
\end{figure*}

\paragraph{Fixpoints}

For long time scales, we observe that the activity in the SIRS model
typically does not converge to zero. We here investigate the onset
of this regime using a simplified version of the dynamical equations
\prettyref{eq:SIR_dynamics_MF}, \prettyref{eq:SIR_dynamics_TAP},
where we set the left hand side to zero $\Delta\rho^{\alpha}=0$.
Solutions to these equations are therefore fixpoints of the average
dynamics. Finding a fixpoint where $\rho^{I}\neq0$ corresponds to
a state with finite activity in the system at all times. We may then
drop the time dependence, since by construction solutions to the fixpoint
equations no longer evolve. We further propose a solution where all
nodes have equal probability of being in state $\alpha$, 
\begin{equation}
\rho_{i}^{\alpha}=\bar{\rho}^{\alpha}\quad\forall i\,.\label{eq:homogeneity_assumption}
\end{equation}
This solution should be attained only in networks which are sufficiently
homogeneous. In random regular networks, although they are not as
homogeneous as, e.g., a 2d square grid, all nodes have equal degree.
Thus, under the assumption that \prettyref{eq:homogeneity_assumption}
holds, sums over nearest neighbors of nodes simplify, $\sum_{j}a_{ij}\rho_{j}^{I}\rightarrow k_{i}\bar{\rho}^{I},$with
$k_{i}$ the degree of the $i$-th node, which is equal to $k$ for
all $i$ in random regular networks. The mean-field solution to the
homogeneous fixpoint-equations is then  
\begin{equation}
\bar{\rho}_{\text{MF}}^{I}\in\left\{ 0,\frac{\eta}{\eta+\mu}\left(1-\frac{1}{\lambda k}\right)\right\} .\label{eq:rho_bar_I_MF}
\end{equation}
 The trivial solution, $\bar{\rho}_{\text{MF}}^{I}=0$, always exists.
Since we must have $\bar{\rho}_{\text{MF}}^{I}\geq0$, the non-trivial
solution emerges as soon as $\lambda\ge\lambda_{c}^{\text{MF}}:=k^{-1}.$

To derive a similar prediction for the TAP equations, we introduce
another dynamical variable $\tau_{i}(t)=\sum_{j}a_{ij}\rho_{j\text{\ensuremath{\leftarrow i}}}^{I}(t)$.
We then find 
\begin{align}
\Delta\tau_{i}(t+1) & =-\mu\tau_{i}(t)+\beta\sum_{j}a_{ij}a_{ji}\rho_{j}^{S}(t)\rho_{i}^{I}(t)\label{eq:def_tau}
\end{align}
and we can replace the term $\beta\sum_{j}a_{ij}\rho_{j\leftarrow i}(t)$
by $\tau_{i}(t)$ in \prettyref{eq:SIR_dynamics_TAP}. An explicit
statement of all evolution equations using $\tau$ is given in \prettyref{app:overview}.
We now make the same homogeneity assumption, set $\Delta\overline{\rho}^{\alpha}=0,\:\Delta\overline{\tau}=0$
and drop the time arguments. This yields the conditions
\begin{align*}
\rho^{R} & =\frac{\mu}{\eta}\bar{\rho}^{I}\\
0 & =-\mu\bar{\rho}^{I}+\bar{\rho}^{S}\beta k\bar{\rho}^{I}-\bar{\rho}^{S}\beta\tau\\
0 & =-\mu\tau+\bar{\rho}^{I}\beta k\bar{\rho}^{S}\,.
\end{align*}
Thus, we obtain a cubic equation in $\bar{\rho}^{S}$ with the solutions
\[
\bar{\rho}_{\text{TAP}}^{S}\in\left\{ 1,\frac{1}{\text{\ensuremath{2\lambda}}}\left(1\pm\sqrt{1-\frac{4}{k}}\right)\right\} ,
\]
For $k<4$, we find only the trivial solution $\bar{\rho}^{S}=1.$
For $k\geq4$ , a non-trivial solution emerges as soon as $\lambda\ge\lambda_{c}^{\text{TAP}}:=\frac{1}{\text{\ensuremath{2}}}\left(1-\sqrt{1-\frac{4}{k}}\right)$.
We here choose the solution with the minus sign as this yields the
lowest $\lambda$ at which $\bar{\rho}_{\text{TAP}}^{S}<1$ holds.
. Note that for $k\gg4$, this again reproduces the mean-field result
$\lambda_{c}^{\text{TAP}}\approx k^{-1}+\mathcal{O}\left(\left(\frac{4}{k}\right)^{2}\right).$
The corresponding curves for $\overline{\rho}^{I}$ read 
\begin{equation}
\overline{\rho}_{\text{\text{TAP}}}^{I}\in\left\{ 0,\frac{\eta}{\eta+\mu}\left(1-\frac{1}{\text{\ensuremath{2\lambda}}}\left(1\mp\sqrt{1-\frac{4}{k}}\right)\right)\right\} \,.\label{eq:rho_bar_I_TAP}
\end{equation}
it is interesting to observe that the curve for $\bar{\rho}_{\text{MF}}^{I}$
for $k=2$ thus coincides with the corresponding curve $\overline{\rho}_{\text{\text{TAP}}}^{I}$
for $k=4$. This means that the TAP prediction yields the same endemic
state, at double connectivity for this special configuration. This
illustrates the impact sparsity on the dynamics. 

We compare these predictions to simulations on random regular graphs
in \prettyref{fig:SIRS_rhoI_endemic}. First, we verify that the simplified
Eqs. \eqref{eq:rho_bar_I_MF}, \eqref{eq:rho_bar_I_TAP} indeed match
the outcome of the full update Eqs. \eqref{eq:SIR_dynamics_MF} and
\eqref{eq:SIR_dynamics_TAP}. This is shown in \prettyref{fig:SIRS_rhoI_endemic}(a).
We also find that the mean-field prediction for $k=2$ indeed coincides
with the corresponding curve for the TAP prediction at $k=4$.

We then average the outcome of the simulation for $\rho^{I}$ over
the last $500$ time steps, and compare this result to mean-field
and TAP theory for varying $\lambda$ and $k$ in \prettyref{fig:SIRS_rhoI_endemic}
(b) and (c). Again, we find a better agreement between TAP theory
and the simulation. In particular, TAP theory accurately predicts
vanishing activity at $k<4$. However, TAP theory still overestimates
the average activity slightly both in the dynamics (see e.g. \prettyref{fig: SIRS_slowly_Waning}(c))
and in the endemic state (e.g. predicting a non-vanishing contribution
at $\lambda\approx0.7$ and $k=4$ in \prettyref{fig:SIRS_rhoI_endemic}(b)).
If the mechanism which leads to the deviation between the fluctuation
correction and the simulation were the omission of the positive self-feedback
loop which we described at the outset of this section, then the fraction
of infected agents in the endemic state should be underestimated by
the fluctuation correction. Contrary to this, we rather find that
it is overestimated by the fluctuation correction.

\section{Discussion\label{sec:Discussion}}

We have demonstrated that a fluctuation correction to the SIR equations
cancels a spurious self-feedback effect, which leads to inflated infection
curves in the standard mean-field version of the SIR model. As a
result, we find that the updated model not only correctly predicts
a significantly lower peak of infection, but also a far larger fraction
of susceptible\textit{ and therefore never infected} individuals in
the final endemic state. This shows that there exists an important
distinction between the stochastic process  and the mean-field approximation
of the former process. Classical SIR models hence represent a different
process, in which self-feedback is present.

Note that the correction term in \prettyref{eq:SIR_dynamics_TAP}
only cancels self-feedback up to second order in $\beta$. The expansion
outlined in \prettyref{app:Expansion_of_effective_action} can be
extended to higher order corrections, which may also include higher
order effects such as self-feedback effects along loops of the network,
e.g. contributions $\propto\beta^{3}a_{ij}a_{jk}a_{ki}$ , corresponding
to a self-feedback term along the loop from $i$ to $j$ to $k$ and
then back to $i$. This may be relevant for network topologies with
large clustering coefficients. Furthermore, we have assumed the network
connectivity to be constant; however the theoretical framework allows
for a time dependence of the connectivity matrix as well. We leave
an investigation of higher order corrections and network temporal
variations of the connectivity to future work.

Numerous studies have previously demonstrated that network heterogeneity
plays an important role in the predictions of the SIR model \citep{thurner_network-based_2020,heltberg_spatial_2022,tkachenko_time-dependent_2021,newman_spread_2002,moreno_epidemic_2002,may_infection_2001}.
In particular, Thurner et al. \citep{thurner_network-based_2020}
reported a large difference between the completely homogeneous SIR
model and a simulation of the stochastic process, which they attribute
to the network topology. The correction effect we describe here is
relevant for all network topologies, but increasingly so for sparser
networks. In particular, the correction for self-feedback shows that
reducing the number of contacts, thus causing sparser network topology,
has a stronger effect than standard SIR calculations suggest, and
is therefore an even more effective measure to suppress the spread
of disease than previously assumed.

Other than the mean-field approximation both the pair approximation
\citep{mata_pair_2013} and dynamical message passing (DMP) \citep{karrer_message_2010,lokhov_inferring_2014}
have emerged as approximations of the stochastic process. Within the
pair-approximation \citep{henkel_non-equilibrium_2008,ben-avraham_mean-field_1992},
not to be confused with the ``independent pair approximation'',
see e.g. \citep{Roudi09a}, in addition to the first moments, one
also takes pairwise correlations between agents into account. However,
the evolution of first and second moments of the distribution is not
closed, to compute their time evolution, one needs to compute third
moments, which are then approximated using second and first moments.
While the pair approximation certainly improves upon the mean-field
equations, it is less computationally efficient, requiring the tracking
of $\mathcal{O}(N^{2})$ coupled differential equations. In contrast,
in our framework we must only track $\mathcal{O}(N)$ dynamical variables.

Within the SIR model, self-feedback is forbidden, since the transitions
$I\rightarrow S$ and $R\rightarrow S,I$ never occur. This idea of
the cancellation of self-feedback in the SIR model underlies DMP:
within DMP, one tracks messages, namely the probabilities that an
infection is passed along a given edge of the graph. The DMP equations
follow from a strict prohibition of all self-feedback. The number
of dynamical variables tracked in the DMP approach scales with the
number of edges. For the SIR model, DMP is exact on trees.

Within variants of the SIR model where transitions back to the susceptible
state are allowed, such as the SIS or SIRS model, self-feedback may
indeed occur, see e.g. \citep{pastor-satorras_epidemic_2015,goltsev_localization_2012,castellano_cumulative_2020,castellano_relevance_2018}.
Nevertheless, the authors of \citep{shrestha_message-passing_2015}
applied recurrent DMP equations to the SIRS model, and found that
eliminating self-feedback can lead to improved predictions. While
a positive self-feedback loop is certainly present in these systems
\citep{castellano_relevance_2018}, the fluctuation correction to
the SIRS and SIS models has the same functional form as the one to
the SIR model, hence it cancels a self-feedback effect up to the second
order in the interaction strength. However, while the DMP approach
heuristically imposes the non-existence of self-feedback, here its
cancellation emerges from a systematic expansion. Furthermore, while
the DMP equations of \citep{shrestha_message-passing_2015} eliminate
all self-feedback, our update equations in principle allow for self-feedback
of higher orders.

Within recurrent infection models, the allowed self-feedback loop,
in which an initially infected node recovers and becomes susceptible
again between the original infection and the reception of the self-feedback
signal, comes with a factor $\beta^{2}\mu\eta\ll\beta^{2}$ for the
SIRS model, or a factor $\beta^{2}\bar{\mu}\ll\beta^{2}$ for the
SIS model. In the case in which the loss of immunity is significantly
slower than the spread of infection, one could attempt to argue that
this effect becomes negligible, thus justifying the use of DMP. However,
depending on the network topology, the positive self-feedback effect
exists, and can considerably change the predictions \citep{castellano_relevance_2018}.
A systematic comparison between the pair approximation and the DMP
approach to the SIRS model on scale-free graphs has been provided
in \citep{silva_comparison_2022}. The authors of the latter study
find that neither theory predicts both the endemic activity and the
localization of the activity above the epidemic threshold it accurately.

We found that even in systems where a positive self-feedback is in
principle permitted, fluctuations can partially cancel the self-feedback
effect. This observation can explain why an elimination of all self-feedback
effects, reported in \citep{shrestha_message-passing_2015} via dynamic
message passing, can yield improved results, even in systems where
positive self-feedback loops are allowed. However, our results do
not justify the complete cancellation of all self-feedback, which
is known to lead to incorrect predictions \citep{castellano_relevance_2018}.
Rather, our theory presents a middle ground between a complete elimination
of the self-feedback effect and its over-representation in mean-field
theory.

For the SIR model our result eliminates the strongest self-feedback
effect, while it does not eliminate self-feedback altogether. For
recurrent infection models such as the SIRS model, this partial elimination
must be viewed in a different light: there, it is only a correction
on the level of fluctuations. From the form of this correction term,
we cannot conclude that self-feedback in all forms is absent in SIRS
or SIS models. Nevertheless, it can explain the partial success of
DMP theory also in recurrent infection models, since it illustrates
that fluctuations can lead to a cancellation of self-feedback to a
certain extent. We expect that higher than second order corrections
to the dynamics will differ in their nature depending on which variant
of the three models one chooses.

We have found that both spin systems at equilibrium and disease models,
can be linked via the same expansion method. It is interesting to
observe that additionally,  the expansion produces the cancellation
of self-feedback in both models at second order. We have demonstrated
that the expansion method can be applied also to other variants of
the SIR model. This can be further extended, by incorporating for
example the SEIR or SEIRS model. Moreover, the computation of higher-order
corrections is straightforward. 

Overall, it is usually possible to argue whether self-feedback should
be present or absent in a system without a systematic fluctuation
expansion, and the presence or absence of the same effect can have
dramatic effects on the system's behavior. However, such a heuristic
argument is not always sufficient to predict properties of the system
as we saw in SIRS model. In particular, the non-cancellation of self-feedback
does not imply that a first order mean-field theory is accurate. In
these cases, a systematic fluctuation expansion can yield improved
predictions. Computing these corrections in different systems in which
self-feedback is allowed is an interesting direction of further research.
\begin{acknowledgments}
This work was partly supported by the German Federal Ministry for
Education and Research  (BMBF Grant 01IS19077B to Aachen) and by NeuroSys
as part of the initiative \textquotedblleft Clusters4Future\textquotedblright ,
which is funded by the Federal Ministry of Education and Research
BMBF (03ZU1106CB). It was also partly supported by the European Union\textquoteright s
Horizon 2020 Framework Programme for Research and Innovation under
Specific Grant Agreement No. 945539 (Human Brain Project SGA3). We
acknowledge funding by the Deutscher Akademischer Austauschdienst
- DAAD scholarship programme \textquotedbl Forschungsstipendien für
Doktorandinnen und Doktoranden 2022/2023\textquotedbl{} (DAAD, funding
programme 57595573).
\end{acknowledgments}

\appendix

\onecolumngrid

\section{Evolution equations for SIR, SIRS, and SIS model\label{app:overview}}

In \prettyref{tab:Overview_table}, we present the corrected evolution
equations for all three models seperately. Analogous to \prettyref{sec:Experiments},
we will introduce another dynamical variable $\tau_{i}$ to account
for the TAP term; this makes it possible to implement the TAP evolution
equations as $3N$ coupled update equations. 
\begin{table}
\begin{centering}
\begin{tabular}{|c|c|}
\hline 
model & evolution equations\tabularnewline
\hline 
 & $\Delta\rho_{i}^{S}(t+1)=-\rho_{i}^{S}(t)\beta\left(\sum_{j}a_{ij}\rho_{j}^{I}(t)-\tau_{i}(t)\right)$\tabularnewline
SIR & $\Delta\rho_{i}^{I}(t+1)=-\mu\rho_{i}^{I}(t)+\rho_{i}^{S}(t)\beta\left(\sum_{j}a_{ij}\rho_{j}^{I}(t)-\tau_{i}(t)\right)\,,$\tabularnewline
 & $\Delta\tau_{i}(t+1)=(1-\mu)\tau_{i}(t)+\beta\sum_{j}a_{ji}\rho_{j}^{S}(t)\rho_{i}^{I}(t)$\tabularnewline
\hline 
 & $\Delta\rho_{i}^{S}(t+1)=\bar{\mu}\rho_{i}^{I}(t)-\rho_{i}^{S}(t)\beta\left(\sum_{j}a_{ij}\rho_{j}^{I}(t)-\tau_{i}(t)\right)$\tabularnewline
SIS & $\Delta\rho_{i}^{I}(t+1)=-\bar{\mu}\rho_{i}^{I}(t)+\rho_{i}^{S}(t)\beta\left(\sum_{j}a_{ij}\rho_{j}^{I}(t)-\tau_{i}(t)\right)\,,$\tabularnewline
 & $\Delta\tau_{i}(t+1)=(1-\bar{\mu})\tau_{i}(t)+\beta\sum_{j}a_{ji}\rho_{j}^{S}(t)\rho_{i}^{I}(t)$\tabularnewline
\hline 
 & $\Delta\rho_{i}^{S}(t+1)=\eta\left(1-\rho_{i}^{I}(t)-\rho_{i}^{S}(t)\right)-\rho_{i}^{S}(t)\beta\left(\sum_{j}a_{ij}\rho_{j}^{I}(t)-\tau_{i}(t)\right)$\tabularnewline
SIRS & $\Delta\rho_{i}^{I}(t+1)=-\mu\rho_{i}^{I}(t)+\rho_{i}^{S}(t)\beta\left(\sum_{j}a_{ij}\rho_{j}^{I}(t)-\tau_{i}(t)\right)\,,$\tabularnewline
 & $\Delta\tau_{i}(t+1)=(1-\mu)\tau_{i}(t)+\beta\sum_{j}a_{ji}\rho_{j}^{S}(t)\rho_{i}^{I}(t)$\tabularnewline
\hline 
\end{tabular}
\par\end{centering}
\caption{Update equations of the TAP theory for the three models. }
\label{tab:Overview_table}
\end{table}

\section{Expansion of the effective action for quenched disorder \label{app:Expansion_of_effective_action}}

We perform the fluctuation expansion analogous to \citep{roudi_dynamical_2011},
with an adapted transition probability. We use a short-hand notation
\[
X_{i}(t)=\left(\begin{array}{c}
S_{i}(t)\\
I_{i}(t)
\end{array}\right)\in\left\{ \left(\begin{array}{c}
1\\
0
\end{array}\right),\left(\begin{array}{c}
0\\
1
\end{array}\right),\left(\begin{array}{c}
0\\
0
\end{array}\right)\right\} 
\]
to describe the state of each individual, and $X(t)$ to indicate
the state of the whole system of nodes at each time point. At each
time step, for each node we hence have a transition probability from
$S_{i}(t),I_{i}(t)$ to $S_{i}(t+1),I_{i}(t+1)$
\begin{align}
W_{t+1,t}\left[X(t+1)\left|\theta(t),X(t)\right.\right]=\prod_{i=1}^{N} & W_{t+1,t}\left[X_{i}(t+1)\left|\theta_{i}(t),X_{i}(t)\right.\right]\nonumber \\
=\prod_{i=1}^{N} & \left\{ S_{i}(t+1)\left[1-\phi\left(\theta_{i}(t)\right)\right]S_{i}(t)\right.\\
 & +I_{i}(t+1)\left[(1-\mu)I_{i}(t)+\phi\left(\theta_{i}(t)\right)S_{i}(t)\right]\nonumber \\
 & \left.+(1-S_{i}(t+1)-I_{i}(t+1))\left[1-S_{i}(t)-(1-\mu)I_{i}(t)\right]\right\} \,.\label{eq:TransitionProbability}
\end{align}
The first line corresponds to the individual remaining susceptible.
The second corresponds to a new or sustained infection. The last line
corresponds to the probability that the individual enters or remains
in the recovered state. We start by defining a cumulant generating
function 
\begin{equation}
\mathcal{W}(\psi,h)=\ln\left\langle \exp\left(\psi^{\T}X\right)\right\rangle \label{eq:Cumulant_generating_function}
\end{equation}
which couples a set of sources 
\[
\psi_{i}(t)=\left(\begin{array}{c}
\psi_{i}^{S}(t)\\
\psi_{i}^{I}(t)
\end{array}\right)
\]
to the fields, where we used the shorthand notation $\psi^{\T}X=\sum_{i,t}\psi_{i}^{\T}(t)X_{i}(t)$.
The right hand side \prettyref{eq:Cumulant_generating_function} depends
on the external fields $h$, since this alters the update probability
$W_{t+1,t}\left[X(t+1)\left|\theta(t),X(t)\right.\right]$ via $\theta$,
see \prettyref{eq:input_field_theta}.

One can compute the cumulants of the vectors $X_{i}(t)$, by taking
the derivatives of \prettyref{eq:Cumulant_generating_function} by
the sources $\psi$ and subsequently setting $\psi=0$. For example,
we may obtain $\rho_{i}^{\alpha}$ by computing

\begin{equation}
\rho_{i}^{\alpha}(t)=\langle X_{i}^{\alpha}(t)\rangle=\left.\partial_{\psi_{i}^{\alpha}(t)}\mathcal{W}(\psi,h)\right|_{\psi=0}\,.\label{eq:average_from_cumulant_gen_function}
\end{equation}
We may express the average in \prettyref{eq:Cumulant_generating_function},
by enforcing the definition of the input fields $\theta$ via a Dirac
delta distribution and using that, conditioned on the input fields
$\theta(t)$, the transitions from $t$ to $t+1$ are independent
across nodes
\begin{align}
\Big\langle\exp\left(\psi^{\T}X\right)\rangle_{h}=\,\sum_{X}p(X(0))\prod_{i,t}\,\int\frac{d\theta_{i}(t)d\hat{\theta}_{i}(t)}{2\pi} & \exp\left(\psi_{i}^{\T}(t)X_{i}(t)\right)\nonumber \\
\cdot & W_{t+1,t}\left[X_{i}(t+1)\left|\theta_{i}(t),X_{i}(t)\right.\right]\,\nonumber \\
\cdot & \exp\left(i\hat{\theta}_{i}(t)\left[\theta_{i}(t)-h_{i}(t)-\beta\sum_{j}a_{ij}I_{j}(t)\right]\right)\label{eq:Z_explicit}
\end{align}
Here, we introduced an auxiliary field $\hat{\theta}_{i}(t)$ to enforce
the condition \prettyref{eq:input_field_theta} on $\theta_{i}(t)$
via the inverse Fourier transform $\delta(x)=\int\frac{d\hat{x}}{2\pi}\exp(i\hat{x}x)$.
The sum $\sum_{X}$ runs over all trajectories, i.e. all combinations
of all configurations $X$ across time $t\in\left\{ 0,\dots,T\right\} $.
We must also average over the starting density $p(X(0))$, which we
assume to be known. Writing \prettyref{eq:Z_explicit} in this way,
we can effectively split the role of the input fields $\theta$ from
those of the binary variables $X(t)$. We can think of \prettyref{eq:Z_explicit}
as a reordering of the averages: starting with a known density $p(X(0))$,
we compute the sum over the $X(0),$ which gives us a statistic of
the $\theta(0)$ variables. We then integrate over $\theta(0),\hat{\theta}(0)$
to get the statistic of the $X(1)$, and then again sum over the $X(1)$
to get the statistic of the $\theta(1),\hat{\theta}(1),$and so on.
In practice however, these computations are not done exactly. In the
following, we will specify the expansion which allows us to compute
averages $\rho^{\alpha}(t)$ perturbatively.

First, observe that up to a prefactor $-i$, $\hat{\theta}_{i}(t)$
couples to $h_{i}(t)$ in the same manner as the source fields $\psi$
couple to the physical observables $X$. We hence introduce another
average
\begin{equation}
\rho_{\hat{\theta_{i}}}(t)=\langle\hat{\theta}_{i}(t)\rangle=-i\left.\partial_{h_{i}(t)}W(\psi,h)\right|_{\psi=0}\,=0\,,\label{eq:auxiliary_average}
\end{equation}
which vanishes due to the normalization condition on $W(\psi,h)$.
We then use \prettyref{eq:average_from_cumulant_gen_function} and
\prettyref{eq:auxiliary_average} to define the Legendre-Fenchel transform
of \prettyref{eq:Cumulant_generating_function}, which we will refer
to as the effective action, via
\begin{equation}
\Gamma\left(\rho,\rho_{\hat{\theta}}\right)=\sup_{\psi,h}\psi^{\T}\rho-ih^{\T}\rho_{\hat{\theta}}-\mathcal{W}(\psi,h)\,.\label{eq:Gamma}
\end{equation}
Together with \prettyref{eq:average_from_cumulant_gen_function} and
\prettyref{eq:auxiliary_average}, this yields the equations of state
\begin{align}
\partial_{\rho_{i}^{\alpha}(t)}\Gamma\left(\rho,\rho_{\hat{\theta}}\right) & =\psi_{i}^{\alpha}(t)\nonumber \\
\partial_{\rho_{\hat{\theta_{i}}}(t)}\Gamma\left(\rho,\rho_{\hat{\theta}}\right) & =-ih_{i}(t)\,.\label{eq:eq_of_state}
\end{align}
To obtain the correction to the mean-field theory, we then proceed
to expand \prettyref{eq:Gamma} to second order in $\beta$ around
the non-interacting case $\beta=0$, 
\begin{equation}
\text{\ensuremath{\Gamma\approx\Gamma_{\beta=0}+\beta\partial_{\beta}\Gamma+\frac{1}{2}\beta^{2}\partial_{\beta}^{2}\Gamma\,.}}\label{eq:Gamma_expansion}
\end{equation}
 Reinserting this into \prettyref{eq:eq_of_state}, to first order
in $\beta$, one obtains the mean-field approximation \prettyref{eq:SIR_dynamics_MF},
and the dynamical TAP equation \prettyref{eq:SIR_dynamics_TAP} to
second order.

We will now perform the expansion of $\Gamma$ to second order in
$\beta$ term by term.

\subsection{Noninteracting system }

At $\beta=0,$ the cumulant generating function decomposes into a
sum, since $X_{i},X_{j}$ are now independent for $i\neq j$, 
\begin{equation}
\mathcal{W}_{\beta=0}(\psi,h)=\sum_{i}\ln Z_{i}(\psi_{i},h_{i})\,,\label{eq:cum_gen}
\end{equation}
we can view the terms $Z_{i}(t)$ as single-agent partition functions.

We must find \prettyref{eq:Gamma} under the conditions \eqref{eq:auxiliary_average},
and \eqref{eq:average_from_cumulant_gen_function}. To this end, we
must express the fields $\psi,h$ as functions of the mean values
$\rho_{i}^{\alpha},\rho_{\hat{\theta}}$. In the non-interacting case,
the full information about the distribution of each of the binary
variables at any time point is contained in their mean. This is so,
because the variables are independent and binary, so that the individual
means determine the entire distribution. The Markov property of the
random process, the property that the present time step depends only
on the previous one and not on its history, neither on future time
steps further allows us to write down the moment generating function
of a single time step from $t$ to $t+1$ with the averages at $t$
given. We then have
\begin{align*}
Z_{i}(\psi_{i},h_{i})= & e^{\psi_{i}^{S}(t+1)}\underbrace{\rho_{i}^{S}(t)\left[1-\phi(h_{i}(t))\right]}_{S\rightarrow S}\\
 & +e^{\psi_{i}^{I}(t+1)}\left[\underbrace{\phi(h_{i}(t))\rho_{i}^{S}(t)}_{S\rightarrow I}+\underbrace{(1-\mu)\rho_{i}^{I}(t)}_{I\rightarrow I}\right]\\
 & +\underbrace{\mu\rho_{i}^{I}(t)}_{I\rightarrow R}+\underbrace{1-\rho_{i}^{S}(t)-\rho_{i}^{I}(t)}_{R\rightarrow R}.
\end{align*}
We have labeled the transition probabilities by the transitions in
brackets, e.g. $S\rightarrow I$ for the transition from a susceptible
to infected state. We must now find \prettyref{eq:Gamma} under the
conditions \prettyref{eq:auxiliary_average}, \prettyref{eq:average_from_cumulant_gen_function}.
We first evaluate \prettyref{eq:auxiliary_average} from \prettyref{eq:cum_gen}
\[
\rho_{\hat{\theta_{i}}(t)}=\frac{-i\partial_{h_{i}(t)}Z_{i}(\psi_{i},h_{i})}{Z_{i}(\psi_{i},h_{i})}=\frac{-i\phi^{\prime}(h_{i}(t))\rho_{i}^{S}(t)\left(e^{\psi_{i}^{I}(t+1)}-e^{\psi_{i}^{S}(t+1)}\right)}{Z_{i}(\psi_{i},h_{i})}=0\,,
\]
hence we find for $\phi^{\prime}(h_{i}(t))\rho_{i}^{S}(t)\neq0$ that
$\psi_{i}^{I}(t)=\psi_{i}^{S}(t)\,\forall i,t$. The derivatives by
the sources $\psi^{S}$ and $\psi^{I}$ yield
\begin{align*}
\rho_{i}^{S}(t+1)= & \frac{e^{\psi_{i}^{S}(t+1)}\left[1-\phi(h_{i}(t))\right]\rho_{i}^{S}(t)}{Z_{i}(t)}\\
\rho_{i}^{I}(t+1)= & \frac{e^{\psi_{i}^{I}(t+1)}\left(\phi(h_{i}(t))\rho_{i}^{S}(t)+(1-\mu)\rho_{i}^{I}(t)\right)}{Z_{i}(t)}
\end{align*}
we solve these equations for $\psi_{i}^{S}(t),h_{i}(t)$, and obtain
\begin{align*}
\psi_{i}^{S}(t) & =\ln\frac{\rho_{i}^{S}(t)+\rho_{i}^{I}(t)}{1-\rho_{i}^{S}(t)-\rho_{i}^{I}(t)}-\ln\frac{\rho_{i}^{S}(t-1)+(1-\mu)\rho_{i}^{I}(t-1)}{1-\rho_{i}^{S}(t-1)-(1-\mu)\rho_{i}^{I}(t-1)}\\
h_{i}(t) & =\phi^{-1}\left(\frac{\rho_{i}^{I}(t+1)\rho_{i}^{S}(t)-\rho_{i}^{S}(t+1)(1-\mu)\rho_{i}^{I}(t)}{\rho_{i}^{S}(t+1)\rho_{i}^{S}(t)+\rho_{i}^{I}(t+1)\rho_{i}^{S}(t)}\right)\,.
\end{align*}
We insert this back into $Z$, to obtain for the partition function
\[
Z_{i}(\psi_{i},h_{i})=\prod_{t}\,\frac{\left(1-\rho_{i}^{S}(t)-(1-\mu)\rho_{i}^{I}(t)\right)}{1-\rho_{i}^{S}(t+1)-\rho_{i}^{I}(t+1)}\,.
\]
This quantity must equal one, since the partition function is normalized.
This equation then simply expresses that the probability $1-\rho_{i}^{S}(t)-\rho_{i}^{I}(t)$
that node $i$ is recovered, increases by $\mu\rho_{i}^{I}(t)$ in
each time step.

\subsection{Mean-field\label{subsec:Mean-field-SIR} }

We will now compute the first order correction to $\Gamma$. To do
so, we write 
\begin{align*}
\Gamma & =\ln\text{\ensuremath{\sum_{X}\rho\left(X(0)\right)}}\prod_{i,t}\,\int\frac{d\theta_{i}(t)d\hat{\theta}_{i}(t)}{2\pi}\,\exp(\Omega_{\beta})W_{t+1,t}\left[X(t+1)\left|\theta(t),X(t)\right.\right]
\end{align*}
with $\Omega_{\beta}$ defined by 
\begin{align}
\Omega_{\beta}=\sum_{i,t}\Bigg[ & \psi_{i}(t)^{\T}\left(\rho_{i}(t)-X_{i}(t)\right)+ih_{i}(t)\left(\hat{\theta_{i}}(t)-\rho_{\hat{\theta_{i}}}(t)\right)\nonumber \\
 & -i\hat{\theta}_{i}(t)\left(\theta_{i}(t)-\beta\sum_{j}a_{ij}I_{j}(t)\right)\,\Bigg].\label{eq:Omega}
\end{align}
We may now use that, similar to the cumulant generating function,
the derivative yields another average $\partial_{\beta}\Gamma=\langle\partial_{\beta}\Omega_{\beta}\rangle$.
In the case that $\beta=0$, all averages belonging to different indices
factorize, because nodes $i,j$ are independent for $i\neq j$. We
hence find that 
\begin{align*}
\left.\partial_{\beta}\Gamma\right|_{\beta=0} & =\left.\langle\partial_{\beta}\Omega_{\beta}\rangle\right|_{\beta=0}\\
 & =i\sum_{i,t}\rho_{\hat{\theta_{i}}}(t)\sum_{j}a_{ij}\rho_{j}^{I}(t)\,.
\end{align*}
To first order therefore, the effective action reads
\begin{align}
\Gamma=\sum_{i,t}\Bigg[ & \ln\left(\frac{1-\rho_{i}^{S}(t+1)-\rho_{i}^{I}(t+1)}{\left(1-\rho_{i}^{S}(t)-(1-\mu)\rho_{i}^{I}(t)\right)}\right)\nonumber \\
 & +\left(\rho_{i}^{S}(t)+\rho_{i}^{I}(t)\right)\left(\ln\frac{\rho_{i}^{S}(t)+\rho_{i}^{I}(t)}{1-\rho_{i}^{S}(t)-\rho_{i}^{I}(t)}+\ln\frac{1-\rho_{i}^{S}(t-1)-(1-\mu)\rho_{i}^{I}(t-1)}{\rho_{i}^{S}(t-1)+(1-\mu)\rho_{i}^{I}(t-1)}\right)\nonumber \\
 & -i\rho_{\hat{\theta_{i}}}(t)\phi^{-1}\left(\frac{\rho_{i}^{I}(t+1)\rho_{i}^{S}(t)-\rho_{i}^{S}(t+1)(1-\mu)\rho_{i}^{I}(t)}{\rho_{i}^{S}(t)\left(\rho_{i}^{S}(t+1)+\rho_{i}^{I}(t+1)\right)}\right)\nonumber \\
 & +i\rho_{\hat{\theta_{i}}}(t)\beta\sum_{j}a_{ij}\rho_{j}^{I}(t)\,\Bigg]+\mathcal{O}\left(\beta^{2}\right)\label{eq:Gamma_MF}
\end{align}
 To obtain the mean-field equation, we must take the derivative after
$\rho_{i}(t),\rho_{\hat{\theta_{i}}}(t)$ and set the right hand side
to zero. We find that both identities in \prettyref{eq:eq_of_state}
are solved by 
\[
\rho_{i}^{S}(t)+\rho_{i}^{I}(t)=\rho_{i}^{S}(t-1)+(1-\mu)\rho_{i}^{I}(t-1)\,.
\]
Taking the derivative by $\rho_{\hat{\theta_{i}}}(t)$ then yields
the mean-field equations 
\begin{align*}
\Delta\rho_{i}^{S}(t+1) & =-\rho_{i}^{S}(t)\phi\left(\beta\sum_{j}a_{ij}\rho_{j}^{I}(t)\right)\\
\Delta\rho_{i}^{I}(t+1)= & -\mu\rho_{i}^{I}(t)+\rho_{i}^{S}(t)\phi\left(\beta\sum_{j}a_{ij}\rho_{j}^{I}(t)\right)\,.
\end{align*}
With $\phi$ the identity, we arrive at \prettyref{eq:SIR_dynamics_MF}.

\subsection{Second order correction\label{app:Second-order-correction}}

We compute the second derivative of $\Gamma$ 
\begin{equation}
\partial_{\beta}^{2}\Gamma_{\beta=0}=\Big\langle\partial_{\beta}^{2}\Omega_{\beta}\Big\rangle_{\beta=0}+\Big\langle\left(\partial_{\beta}\Omega_{\beta}\right)^{2}\Big\rangle_{\beta=0}-\Big\langle\partial_{\beta}\Omega_{\beta}\Big\rangle_{\beta=0}^{2}\,.\label{eq:Gamma_second_order_deriv}
\end{equation}
The first term vanishes. We are left with the variance of $\partial_{\beta}\Omega_{\beta}$
in the non-interacting case. Since we are at second order, the external
fields acquire a linear dependence on $\beta$ via their dependence
on expectation values $\rho_{i}^{\alpha},\rho_{\hat{\theta}}$. We
must therefore consider derivatives of the type $\partial_{\beta}\psi_{i}^{\alpha}(t)=\partial_{\beta}\partial_{\rho_{i}^{\alpha}(t)}\Gamma$
and $\partial_{\beta}h_{i}(t)=i\partial_{\beta}\partial_{\rho_{\hat{\theta_{i}}}(t)}\Gamma$:
\begin{align}
\partial_{\beta}\Omega_{\beta} & -\left\langle \partial_{\beta}\Omega_{\beta}\right\rangle _{\beta=0}=i\sum_{i,j,t}a_{ij}\delta\hat{\theta}_{i}(t)\delta I_{j}(t)\,,\label{eq:Omega_minus_Omega_exp}
\end{align}
where we define $\delta\hat{\theta}_{i}(t)=\hat{\theta}_{i}(t)-\rho_{\hat{\theta}_{i}}(t)$
and $\delta I_{j}(t)=I_{j}(t)-\rho_{j}^{I}(t)$. Evaluating the averages
at $\beta=0$ again means that all nodes of unequal indices can be
treated independently. In the following, we will always evaluate expectation
values in the non-interacting case and drop the subscript $\beta=0$
for brevity. We hence arrive at
\begin{equation}
\partial_{\beta}^{2}\Gamma_{\beta=0}=-\sum_{t,t^{\prime}}\sum_{ijkl}a_{ij}a_{kl}\left\langle \delta\hat{\theta}_{i}(t)\delta\hat{\theta}_{k}(t^{\prime})\delta I_{j}(t)\delta I_{l}(t^{\prime})\right\rangle \,.\label{eq:Gamma_second_order_deriv_quenched}
\end{equation}
If all indices $i,\dots,l$ are unequal, then the term under the sum
vanishes, since all nodes decouple. We must therefore have at least
two indices equal to get a meaningful contribution.

We now go through the combinations of indices to determine which yield
a meaningful contribution. We immediately observe that due to the
prefactor of $a_{ij}a_{kl}$, we must have that $i\neq j$ and $k\neq l$.
Due to this, the average in \prettyref{eq:Gamma_second_order_deriv_quenched}
must always decompose into at least two factors. Furthermore, in each
factor, each index must be equal to at least one other index, otherwise
the term is proportional to $\langle\delta\hat{\theta}\rangle=0$
or $\langle\delta I\rangle=0$. From the latter two observations it
follows that exactly two independent indices are left, thus the average
in \prettyref{eq:Gamma_second_order_deriv_quenched} composes into
exactly two factors. Thus we must either pair a factor of $\delta\hat{\theta}$
with another factor $\delta\hat{\theta}$ with the same node index,
or a pair a factor of $\delta\hat{\theta}$ with another factor $\delta I$
with the same node index. From 
\begin{equation}
\left\langle \delta\hat{\theta}_{i}(t)\delta\hat{\theta}_{i}(t^{\prime})\right\rangle =0\quad\forall i,t,t^{\prime}
\end{equation}
it follows that only the term where $i=l$ and $k=j$ remains, where
each factor of $\delta\hat{\theta}$ is paired another factor $\delta I$
with the same node index. 

All in all, we find 
\begin{align}
\partial_{\beta}^{2}\Gamma_{\beta=0}= & -\sum_{t,t^{\prime}}\sum_{ij}a_{ij}a_{ji}\left\langle \delta\hat{\theta}_{i}(t)\delta I_{i}(t^{\prime})\right\rangle \left\langle \delta\hat{\theta}_{j}(t^{\prime})\delta I_{j}(t)\right\rangle .\label{eq:Gamma_second_deriv_beta}
\end{align}

We must therefore compute expectation values of the form $\langle\hat{\theta}_{i}(t)I_{i}(t^{\prime})\rangle$.
We will see in the following that $\langle\hat{\theta}_{i}(t)I_{i}(t^{\prime})\rangle$
vanishes unless $t<t^{\prime}$. This is so, because $\langle\hat{\theta}_{i}(t)I_{i}(t^{\prime})\rangle$
has the role of a response function: it measures the effect of a change
in the field $h_{i}(t)$ at time $t$ on the random variable $I_{i}(t^{\prime})$
at another time point. In the case $t\geq t^{\prime}$, therefore,
we find that the factor in \prettyref{eq:Gamma_second_order_deriv}
simplifies to
\begin{align*}
\left\langle \delta\hat{\theta}_{i}(t)\delta I_{i}(t^{\prime})\right\rangle  & =\left\langle \hat{\theta}_{i}(t)I_{i}(t^{\prime})\right\rangle -\rho_{\hat{\theta}_{i}}(t)\rho_{i}^{I}(t^{\prime})\\
 & \overset{t\geq t^{\prime}}{=}-\rho_{\hat{\theta}_{i}}(t)\rho_{i}^{I}(t^{\prime})\,.
\end{align*}
Since the stochastic process is causal (later changes in the external
field can have no influence on earlier time points), the response
function is causal as well, and vanishes for $t\geq t^{\prime}$.
To evaluate the response function, we only need to consider the generating
function from $t$ to $t^{\prime}$ with the averages at $t-1$ given.
Again, this is because of the Markov property of the random process.

For $\beta=0$, it suffices to write down the moment generating function
for a single node $i$:

\begin{align}
Z_{i}(\psi_{i},h_{i})=\rho_{i}^{S}(t)\Bigg[ & \prod_{\tau=t}^{t^{\prime}-1}\left[1-\phi(h_{i}(\tau))\right]e^{\psi_{i}^{S}(\tau+1)}\nonumber \\
 & +\sum_{t_{\text{inf}}=t}^{t^{\prime}-1}\phi(h_{i}(t_{\text{inf}}))e^{\psi_{i}^{I}(t_{\text{inf}}+1)}\left\{ \prod_{\tau=t}^{t_{\text{inf}}-1}\left[1-\phi(h_{i}(\tau))e^{\psi_{i}^{S}(\tau+1)}\right]\right\} \nonumber \\
 & \cdot\left\{ \prod_{\nu=t_{\text{inf}}+1}^{t^{\prime}-1}\left[(1-\mu)e^{\psi_{i}^{I}(\nu+1)}\right]+\sum_{t_{\text{rec}}=t_{\text{inf}}+1}^{t^{\prime}-1}\mu\prod_{\nu=t_{\text{inf}}+1}^{t_{\text{rec}}}\left[(1-\mu)e^{\psi_{i}^{I}(\nu+1)}\right]\right\} \Bigg]\nonumber \\
+\rho_{i}^{I}(t) & \left\{ \prod_{\nu=t}^{t^{\prime}-1}\left[(1-\mu)e^{\psi_{i}^{I}(\nu+1)}\right]+\sum_{t_{\text{rec}}=t}^{t^{\prime}-1}\mu\prod_{\nu=t}^{t_{\text{rec}}}\left[(1-\mu)e^{\psi_{i}^{I}(\nu+1)}\right]\right\} \nonumber \\
+1-\rho_{i}^{S} & (t)-\rho_{i}^{I}(t)\label{eq:Moment_generating_function}
\end{align}
This sum is organized as follows: The first line corresponds to the
node remaining susceptible. The second line counts all possible times
$t_{\text{inf}}$ of infection. The the first term in the third line
corresponds to a remaining infection until $t^{\prime}$. The second
term counts all possible recovery times respectively. The fourth line
counts all trajectories with initial infection, and respectively the
possibilities of recovery, analogous to the third line. The last line
corresponds to the node beginning in the recovered state. To compute
$\langle\hat{\theta}_{i}(t)I_{i}(t^{\prime})\rangle$, we take the
derivative 
\begin{align}
\langle\hat{\theta}_{i}(t)I_{i}(t^{\prime})\rangle_{\beta=0} & \overset{t^{\prime}>t}{=}\left.\frac{-i\partial_{h_{i}(t)}\partial_{\psi_{i}^{I}(t^{\prime})}Z_{i}(\psi_{i},h_{i})}{Z_{i}(\psi_{i},h_{i})}\right|_{\psi=h=0}\nonumber \\
 & =-i\rho_{i}^{S}(t)(1-\mu)^{t^{\prime}-t-1}\,.\label{eq:Response_func}
\end{align}
Where we used $\phi(0)=0$, $\phi^{\prime}(0)=1$. Observe that the
derivative in \prettyref{eq:Response_func} picks out of all possible
trajectories precisely the one which corresponds an infection of node
$i$ at time point $t$, which lasts until time point $t^{\prime}$
with probability $(1-\mu)^{t^{\prime}-t-1}$. For $t^{\prime}<t$,
no such trajectory exists, hence $\langle\hat{\theta}_{i}(t^{\prime})I_{i}(t)\rangle=0$.
Therefore the product of the two averages always vanishes 
\begin{equation}
\langle\hat{\theta}_{k}(t^{\prime})I_{k}(t)\rangle\langle\hat{\theta}_{i}(t)I_{i}(t^{\prime})\rangle=0\;\forall i,k,t,t^{\prime}\,.\label{eq:Double_response_vanishes}
\end{equation}

Altogether, we have 

\begin{align}
\partial_{\beta}^{2}\Gamma_{\beta=0}=-\sum_{t,t^{\prime}}\sum_{ij}a_{ij}a_{ji}\,\Bigg( & 2i\Theta(t-t^{\prime})\rho_{j}^{S}(t^{\prime})(1-\mu)^{t-t^{\prime}-1}\rho_{\hat{\theta}_{i}}(t)\rho_{i}^{I}(t^{\prime})\nonumber \\
 & +\rho_{\hat{\theta}_{i}}(t)\rho_{i}^{I}(t^{\prime})\rho_{\hat{\theta}_{j}}(t^{\prime})\rho_{i}^{I}(t)\Bigg)\label{eq:Gamma_second_deriv_beta_explicit}
\end{align}
From which we can compute second order correction to \prettyref{eq:Gamma_MF}.
The correction only changes the equation of state originating from
the derivative after $\rho_{\hat{\theta}}(t).$ This is so, because
the factor $\rho_{\hat{\theta}}(t^{\prime})=0$ cancels all contributions
of the correction term to the equation of state. The only relevant
contribution to the equation of state is hence the term linear in
$\rho_{\hat{\theta}}(t).$ All equations of state together finally
yield \prettyref{eq:SIR_dynamics_TAP} .

\section{Extension to SIS and SIRS model\label{app:Extensions-to-SIS-and-SIRS}}

We now extend the calculation to for the SIS and SIRS model. Up to
\eqref{eq:Z_explicit}, the calculations remain unchanged, but now
we must specify a different update probability 
\begin{align*}
W_{t+1,t}\left[X(t+1)\left|\theta(t),X(t)\right.\right] & =\prod_{i=1}^{N}\left\{ S_{i}(t+1)\left(\left[1-\phi\left(\theta_{i}(t)\right)\right]S_{i}(t)+\eta R_{i}(t)+\bar{\mu}I_{i}(t)\right)\right.\\
 & +I_{i}(t+1)\left[(1-\mu-\bar{\mu})I_{i}(t)+\phi\left(\theta_{i}(t)\right)S_{i}(t)\right]\\
 & \left.+R_{i}(t+1)\left[\left(1-\eta\right)R_{i}(t)+\mu I_{i}(t)\right]\right\} \,,
\end{align*}
where now $\bar{\mu}$ is the probability for $I\rightarrow S$ and
$\eta$ for $R\rightarrow S$ and we used $R_{i}(t)=1-S_{i}(t)-I_{i}(t)$.
 We obtain the SIS model by setting $\mu=\eta=0$ . We set $\bar{\mu}=0$
to obtain the the SIRS model.

\subsection{Noninteracting case}

We first compute $\left.\Gamma\right|_{\beta=0}$. We again first
compute the dependence of the sources on the averages. The derivatives
by the sources $\psi^{S}$ and $\psi^{I}$ yield
\begin{align*}
\rho_{i}^{S}(t+1)= & \frac{e^{\psi_{i}^{S}(t+1)}\left[\left(1-\phi(h_{i}(t))-\eta\right)\rho_{i}^{S}(t)+\eta+\left(\bar{\mu}-\eta\right)\rho_{i}^{I}(t)\right]}{Z_{i}(t)}\\
\rho_{i}^{I}(t+1)= & \frac{e^{\psi_{i}^{I}(t+1)}\left(\phi(h_{i}(t))\rho_{i}^{S}(t)+(1-\mu-\bar{\mu})\rho_{i}^{I}(t)\right)}{Z_{i}(t)}\,.
\end{align*}
From the derivative by $h$, we again find that $\psi^{S}$=$\psi^{I}$.
We solve these equations for $\psi_{i}^{S}(t),h_{i}(t)$, and obtain
\begin{align*}
\psi_{i}^{S}(t) & =\ln\frac{\rho_{i}^{S}(t+1)+\rho_{i}^{I}(t+1)}{\rho_{i}^{S}(t)+\rho_{i}^{I}(t)+\eta\left(1-\rho_{i}^{S}(t)-\rho_{i}^{I}(t)\right)-\mu\rho_{i}^{I}(t)}+\ln\frac{\left(1-\eta\right)\left(1-\rho_{i}^{S}(t)-\rho_{i}^{I}(t)\right)+\mu\rho_{i}^{I}(t)}{1-\rho_{i}^{S}(t+1)-\rho_{i}^{I}(t+1)}\\
h_{i}(t) & =\phi^{-1}\left(\frac{-(1-\mu-\bar{\mu})\rho_{i}^{S}(t+1)\rho_{i}^{I}(t)+\rho_{i}^{I}(t+1)\left(\eta+\left(1-\eta\right)\rho_{i}^{S}(t)+\left(\bar{\mu}-\eta\right)\rho_{i}^{I}(t)\right)}{\rho_{i}^{S}(t)\left(\rho_{i}^{S}(t+1)+\rho_{i}^{I}(t+1\right))}\right)\,.
\end{align*}
We insert this back into $Z$, to obtain for the partition function
\[
Z_{i}(\psi_{i},h_{i})=\prod_{t}\,\frac{\left(1-\eta\right)\left(1-\rho_{i}^{S}(t)-\rho_{i}^{I}(t)\right)+\mu\rho_{i}^{I}(t)}{1-\rho_{i}^{S}(t+1)-\rho_{i}^{I}(t+1)}.
\]
This quantity must equal one, since the partition function is normalized.
This equation then simply expresses that the probability $1-\rho_{i}^{S}(t)-\rho_{i}^{I}(t)$
that node $i$ is recovered, changes by $\mu\rho_{i}^{I}(t)-\eta\left(1-\rho_{i}^{S}(t)-\rho_{i}^{I}(t)\right)$
in each time step.

\subsection{Mean-field equations}

The mean-field equations follow analogously to \prettyref{subsec:Mean-field-SIR}.
We first take the derivative
\begin{align*}
\left.\partial_{\beta}\Gamma\right|_{\beta=0} & =\left.\langle\partial_{\beta}\Omega_{\beta}\rangle\right|_{\beta=0}\\
 & =i\sum_{i,t}\rho_{\hat{\theta_{i}}}(t)\sum_{j}a_{ij}\rho_{j}^{I}(t)\,,
\end{align*}
yielding the same contribution to $\Gamma$ as for the SIR model.
To first order therefore, the effective action reads
\begin{align}
\Gamma=\sum_{i,t}\Bigg[ & \ln\left(\,\frac{1-\rho_{i}^{S}(t+1)-\rho_{i}^{I}(t+1)}{\left(1-\eta\right)\left(1-\rho_{i}^{S}(t)-\rho_{i}^{I}(t)\right)+\mu\rho_{i}^{I}(t)}\right)\nonumber \\
 & +\left(\rho_{i}^{S}(t)+\rho_{i}^{I}(t)\right)\left(\ln\frac{\rho_{i}^{S}(t+1)+\rho_{i}^{I}(t+1)}{\rho_{i}^{S}(t)+\rho_{i}^{I}(t)+\eta\left(1-\rho_{i}^{S}(t)-\rho_{i}^{I}(t)\right)-\mu\rho_{i}^{I}(t)}+\ln\frac{\left(1-\eta\right)\left(1-\rho_{i}^{S}(t)-\rho_{i}^{I}(t)\right)+\mu\rho_{i}^{I}(t)}{1-\rho_{i}^{S}(t+1)-\rho_{i}^{I}(t+1)}\right)\nonumber \\
 & -i\rho_{\hat{\theta_{i}}}(t)\phi^{-1}\left(\frac{-(1-\mu-\bar{\mu})\rho_{i}^{S}(t+1)\rho_{i}^{I}(t)+\rho_{i}^{I}(t+1)\left(\eta+\left(1-\eta\right)\rho_{i}^{S}(t)+\left(\bar{\mu}-\eta\right)\rho_{i}^{I}(t)\right)}{\rho_{i}^{S}(t)\left(\rho_{i}^{S}(t+1)+\rho_{i}^{I}(t+1\right)}\right)\nonumber \\
 & +i\rho_{\hat{\theta_{i}}}(t)\beta\sum_{j}a_{ij}\rho_{j}^{I}(t)\,\Bigg]+\mathcal{O}\left(\beta^{2}\right).\label{eq:Gamma_MF-1}
\end{align}
 To obtain the mean-field equation, we must take the derivative after
$\rho_{i}(t),\rho_{\hat{\theta_{i}}}(t)$ and set the right hand side
to zero. We find that both identities in \prettyref{eq:eq_of_state}
are solved by 
\[
\rho_{i}^{S}(t)+\rho_{i}^{I}(t)=\rho_{i}^{S}(t-1)+(1-\mu)\rho_{i}^{I}(t-1)+\eta\left(1-\rho_{i}^{S}(t-1)-\rho_{i}^{I}(t-1)\right)\,.
\]
Taking the derivative by $\rho_{\hat{\theta_{i}}}(t)$ then yields
the mean-field equations 
\begin{align*}
\Delta\rho_{i}^{S}(t+1) & =\eta\left(1-\rho_{i}^{S}(t)-\rho_{i}^{I}(t)\right)+\bar{\mu}\rho_{i}^{I}(t)-\rho_{i}^{S}(t)\phi\left(\beta\sum_{j}a_{ij}\rho_{j}^{I}(t)\right)\\
\Delta\rho_{i}^{I}(t+1)= & -\left(\mu+\bar{\mu}\right)\rho_{i}^{I}(t)+\rho_{i}^{S}(t)\phi\left(\beta\sum_{j}a_{ij}\rho_{j}^{I}(t)\right)\,.
\end{align*}
which now contain the transitions $I\rightarrow S$ and $R\rightarrow S$
with finite probability.

\subsection{Second order correction}

We proceed in the same fashion as in \prettyref{app:Second-order-correction},
and again find that we must compute the moments in \prettyref{eq:Gamma_second_deriv_beta}.
We first compute the correlation $\langle I_{i}(t)I_{i}(t^{\prime})\rangle_{\beta=0}$.
Because we evaluate this at $\beta=0$, this moment corresponds to
a sustained infection between the two time points, which occurs with
probability 
\[
\langle I_{i}(t)I_{i}(t^{\prime})\rangle_{\beta=0}\overset{t^{\prime}\geq t}{=}(1-\mu-\bar{\mu})^{t^{\prime}-t}\rho_{i}^{I}(t)\,,
\]
where now the rate at which the infection decays is $\mu+\bar{\mu}$.
Similarly, the response function is 
\begin{align*}
\langle\hat{\theta}_{i}(t)I_{i}(t^{\prime})\rangle_{\beta=0} & =-i\begin{cases}
\rho_{i}^{S}(t)(1-\mu-\bar{\mu})^{t^{\prime}-t-1} & t^{\prime}>t\\
0 & t^{\prime}<t
\end{cases}\,.
\end{align*}
In principle, trajectories with multiple re-infections between time
points $t$ and $t^{\prime}$ must be taken into account to compute
the average $\langle\hat{\theta}_{i}(t)I_{i}(t^{\prime})\rangle$,
since the transition $I\rightarrow S\rightarrow I$ and $I\rightarrow R\rightarrow S\rightarrow I$
are now allowed. But these terms drop out when we set $\beta=0$.
Finally, we have 
\[
\left\langle \hat{\theta}_{i}(t)\hat{\theta}_{i}(t^{\prime})\right\rangle =0\,,
\]
due to the normalization, as before. We therefore find that the second
order correction yields the same terms as for the SIR model, when
we replace $\mu\rightarrow\mu+\bar{\mu}$. The final update equation
reads (with $\phi$ the identity) 
\begin{align*}
\Delta\rho_{i}^{S}(t+1) & =\eta\left(1-\rho_{i}^{S}(t)-\rho_{i}^{I}(t)\right)+\bar{\mu}\rho_{i}^{I}(t)-\rho_{i}^{S}(t)\beta\sum_{j}a_{ij}\left(\rho_{j}^{I}(t)-\rho_{j\text{\ensuremath{\leftarrow i}}}^{I}(t)\right)\\
\Delta\rho_{i}^{I}(t+1)= & -\left(\mu+\bar{\mu}\right)\rho_{i}^{I}(t)+\rho_{i}^{S}(t)\beta\sum_{j}a_{ij}\left(\rho_{j}^{I}(t)-\rho_{j\text{\ensuremath{\leftarrow i}}}^{I}(t)\right)\,.
\end{align*}
with the self-feedback correction 
\[
\rho_{j\text{\ensuremath{\leftarrow i}}}^{I}(t):=\beta a_{ji}\sum_{t^{\prime}\leq t-1}\rho_{j}^{S}(t^{\prime})\rho_{i}^{I}(t^{\prime})(1-\mu-\bar{\mu})^{t-t^{\prime}-1}\,.
\]
Setting $\bar{\mu}$ to zero, one obtains the equation for the SIRS
model, and setting $\mu,\eta$ to zero, one obtains the update equation
for the SIS model. These equations are stated explicitly in \prettyref{tab:Overview_table}.

\section{Epidemic dynamics in a Spiking Simulator code\label{app:NEST-implementation}}

We implement the dynamics of the SIR, SIRS and SIS model in NEST \citep{gewaltig_nest_2007},
a simulator for spiking neurons. NEST has previously been used to
model binary neurons \citep{Grytskyy13_131}. We here follow the same
approach as in \citep{Grytskyy13_131} to adapt the simulator to the
SIR, SIRS, and SIS dynamics. 

Note that each agent $i$ needs not know the exact state of its nearest
neighbors to compute the update probability, rather, it suffices to
know how many of these nearest neighbors are infected. This information
is encoded in the input $\theta_{i}$ to the agent. We use spikes
to transmit changes to these fields: When an agent $i$ is infected,
it sends out a spike. Upon receiving a spike, the nearest neighbors
$j$ of the agent hence increments their input fields $\theta_{j}$
by $\beta$. When agent $i$ leaves the infected state, it sends two
spikes at the same time. Two simultaneous spikes received by a nearest
neighbor $j$ then result in a reduction of $\theta_{j}$ by $\beta$.
The implementation will be released as open source with one of the
forthcoming releases of the simulator.

\section{Mean-field and TAP solutions\label{app:Mean-field-and-TAP}}

We compute the predictions from mean-field and TAP solutions using
\prettyref{eq:SIR_dynamics_MF} for mean-field and the corresponding
set of coupled differential equations for TAP in \prettyref{tab:Overview_table}.
At the beginning of each simulation, a random subset of agents is
infected, for these we set $\rho_{i}^{S}(0)=0,$ and $\rho_{i}^{I}(0)=1$,
while for all other agents we set $\rho_{i}^{S}(0)=1,$ and $\rho_{i}^{I}(0)=0$.
We also set $\tau_{i}(0)=0$ for all agents for the TAP approximation.
Within the TAP approximation, values of $\theta_{i}(t)$ outside the
$[0,1]$ interval can occur, we hence use a cutoff function $\phi(\theta)=\max(0,\min(1,\theta))$
as indicated in \prettyref{sec:SIR-model} to reset the value of $\theta$
back to the closest physical value. We then compute the average fraction
of individuals in the state $\alpha$ reported in \prettyref{fig: Sparsity_supresses_spread},
\prettyref{fig: SIRS_slowly_Waning} and \prettyref{fig:SIRS_rhoI_endemic}
by summing, $\rho^{\alpha}(t)=\frac{1}{N}\sum_{i}\rho_{i}^{\alpha}(t)$.

\section{Predictions for the epidemic threshold\label{app:Predictions-threshold}}

We will now compute how the endemic state depends on the network topology
and the rates of infection and recovery. The epidemic threshold bears
a resemblance to a phase transition: at a certain parameter $\lambda$,
the system properties fundamentally change: the absorbing state, where
the number of infected agents is zero, becomes unstable. Rather, the
number of infected agents increases. Before we derive the epidemic
threshold from \prettyref{eq:SIR_dynamics_TAP}, we state three known
estimates for the epidemic threshold originating from mean-field theory.
We will here find that all three estimates from the TAP equations
are the same independent of whether one chooses the SIS or SIRS model.

\paragraph{Indiviual-based mean-field theory,}

also referred to as quenched mean-field theory, corresponds to the
evolution equations \eqref{eq:SIR_dynamics_MF} \citep{pastor-satorras_epidemic_2015}.
Linearizing these equations for small initial infection levels yields
the condition 
\begin{equation}
\lambda_{c}^{\mathrm{IBMF}}=\lambda_{a,\mathrm{max}}^{-1}\label{eq:lambda_c_IBMF}
\end{equation}
with $\lambda_{A,\mathrm{max}}$ the largest eigenvalue of the adjacency
matrix. If $\lambda$ is larger than this value, the number of infected
agents will grow according to \prettyref{eq:SIR_dynamics_MF}.

\paragraph{Degree-based mean-field theory,}

which stipulates that all agents of equal degree are statistically
equivalent, yields the following result \citep{PhysRevLett.86.3200}

\begin{equation}
\lambda_{c}^{\mathrm{DBMF}}=\frac{\langle k\rangle}{\langle k^{2}\rangle}\label{eq:lambda_c_DBMF}
\end{equation}
provided that the network has an uncorrelated degree distribution:
the probability $p(k^{\prime}|k)$ that a node of degree $k$ is connected
to a node of degree $k^{\prime}$ is $p(k^{\prime})k^{\prime}/\langle k\rangle$.

\paragraph{DMP}

For $\gamma\leq\frac{5}{2}$, degree-based mean-field theory and individual-based
mean-field theory are expected to yield equivalent results for large
$N$ \citep{pastor-satorras_epidemic_2015}. For $\gamma>\frac{5}{2}$,
individual-based mean-field theory is known to overestimate the epidemic
threshold. This coincides with the point where the largest eigenvalue
of the adjacency matrix localizes due to the self-feedback effect,
see \citep{martin_localization_2014}. The authors of Ref. \citep{martin_localization_2014}
suggest that to assess the relative importance of a node in a graph,
the leading eigenvector of the following matrix $2N\times2B$ should
be considered 
\begin{equation}
M=\begin{pmatrix}a & \mathds{1}-K\\
\mathds{1} & 0
\end{pmatrix}\,,
\end{equation}
where $K_{ij}=\delta_{ij}k_{i}$ is a diagonal matrix of with the
degree $k_{i}$ of the node $i$ on the diagonal. The underlying idea
is that the self-feedback effect makes hubs appear more relevant than
they are. This suggestion coincides with the estimator of the epidemic
threshold for DMP. The estimator for the DMP epidemic threshold is
then the inverse leading eigenvalue of $M$ \citep{shrestha_message-passing_2015,castellano_relevance_2018}.
We will see in the following, that the TAP estimation follows from
a similar $2N\times2N$ matrix, whose off-diagonal blocks however
differ from those in $M$.

\paragraph{TAP prediction.}

We now compute an estimate for $\lambda_{c}$ from \prettyref{eq:SIR_dynamics_TAP}.
To simplify the analysis, we again define dynamical variable $\tau_{i}(t)$
which has the following properties 
\begin{align*}
\tau_{i}(0) & =0\\
\Delta\tau_{i}(t+1) & =-(\mu+\bar{\mu})\tau_{i}(t)+\rho_{i}^{I}(t)\beta\sum_{j}a_{ij}a_{ji}\rho_{j}^{S}(t)\,.
\end{align*}
Note that in comparison to the definition in the main text, \prettyref{eq:def_tau},
$\tau$ here decays with $\mu+\bar{\mu}$, to account for the SIS
($\mu=0$, $\bar{\mu}>0$) and the SIRS ($\bar{\mu}=0$, $\mu>0$)
model at the same time. Using $\tau$, we may write \prettyref{eq:SIR_dynamics_TAP}
as 
\begin{align*}
\Delta\rho_{i}^{I}(t+1)= & -(\mu+\bar{\mu})\rho_{i}^{I}(t)+\rho_{i}^{S}(t)\beta\left(\sum_{j}A_{ij}\rho_{j}^{I}(t)-\tau_{i}(t)\right)\\
\Delta\rho_{i}^{R}(t+1)= & \mu\rho_{i}^{I}(t)-\eta\rho_{i}^{R}(t)\,,
\end{align*}
and we further require that $\rho_{i}^{S}(t)=1-\rho_{i}^{I}(t)$ for
the SIS and $\rho_{i}^{S}(t)=1-\rho_{i}^{R}(t)-\rho_{i}^{I}(t)$ for
the SIRS model. We now examine the regime in which $\rho^{I}$ is
infinitesimally small, hence we keep only terms up to linear order
in $\rho^{I}$. We also set $\rho^{R}$ to zero in the case of the
SIRS model. This can be seen as a small perturbation from the state
of zero activity. If this state is unstable, the number of infected
agents grows, $\Delta\rho^{I}>0$, and we are above the epidemic threshold.
Linearizing in $\rho^{I}$, we find that

\begin{align*}
\rho_{i}^{I}(t)\beta\sum_{j}a_{ij}a_{ji}\rho_{j}^{S}(t) & =\rho_{i}^{I}(t)\beta\sum_{j}a_{ij}a_{ji}\left(1-\rho_{j}^{I}(t)\right)\\
 & =\rho_{i}^{I}(t)\beta\sum_{j}a_{ij}^{2}+\mathcal{O}\left(\left(\rho^{I}(t)\right)^{2}\right)\,,
\end{align*}
where we used that $a=a^{\T}$ in an undirected network. Finally,
we can replace the sum $\sum_{j}a_{ij}^{2}=k_{i}$ by the degree $k_{i}$
of the $i$-th node. The linearized update equations hence read:

\begin{equation}
\begin{pmatrix}\Delta\overrightarrow{\rho^{I}}(t+1)\\
\Delta\overrightarrow{\tau}(t+1)
\end{pmatrix}=\left[-\left(\mu+\bar{\mu}\right)\begin{pmatrix}\mathds{1} & 0\\
0 & \mathds{1}
\end{pmatrix}+\beta\begin{pmatrix}a & -\mathds{1}\\
K & 0
\end{pmatrix}\right]\begin{pmatrix}\overrightarrow{\rho^{I}}(t)\\
\overrightarrow{\tau}(t)
\end{pmatrix}\,,\label{eq:SIRS_linearized}
\end{equation}
with $K_{ij}=\delta_{ij}k_{i}$ a diagonal matrix with the degrees
of the nodes along the diagonal. We now ask that the right hand side
of this equation must be equal to zero: This is exactly the point
where the activity neither decays nor increases and therefore marks
the transition point. This gives us an eigenvalue equation

\begin{equation}
\lambda^{-1}\begin{pmatrix}\rho^{I}(t)\\
\tau(t)
\end{pmatrix}=\underbrace{\begin{pmatrix}a & -\mathds{1}\\
K & 0
\end{pmatrix}}_{=:D}\begin{pmatrix}\rho^{I}(t)\\
\tau(t)
\end{pmatrix}\,,
\end{equation}
where we use $\lambda=\frac{\beta}{\mu}$ in case of the SIS and $\lambda=\frac{\beta}{\bar{\mu}}$
for the SIRS model. The estimator for the epidemic threshold is hence
\begin{equation}
\lambda_{c}^{\text{TAP}}=\left(\lambda_{D}^{\text{max}}\right)^{-1}\,,\label{eq:lambda_c_TAP}
\end{equation}
the inverse leading eigenvalue of $D$. Since the matrix $D$ is not
symmetric, we can obtain complex eigenvalues, corresponding to oscillatory
solutions to \prettyref{eq:SIRS_linearized}.  In the case of random
regular graphs, a homogeneous ansatz for the eigenvector yields $\left(\lambda_{D}^{\text{max}}\right)_{\pm}^{-1}=\frac{1}{2}\left(1\pm\sqrt{1-\frac{4}{k}}\right)$,
consistent with the fixpoint analysis of the previous section. In
the case of scale-free networks, treated in the next section, we found
only real leading eigenvalues of $D$.

\section{Epidemic threshold in scale-free networks\label{app:Epidemic-threshold-scale-free}}

We now move to a different class of networks: scale-free networks.
They are defined by having a degree distribution which follows $p(k)\sim k^{-\gamma}$.
These networks are more heterogeneous: alongside many nodes of low
degree, there are a number of hubs $h$ with degree far above the
network average $k_{h}\gg\langle k\rangle$; the proportion of these
hubs in the population increases for smaller $\gamma$. Scale-free
networks as contact graphs hence represent a more heterogeneous society,
where few individuals have far more contacts than average. In this
section, we present our results on the endemic activity in the SIRS
model in scale-free networks.

We now compare these different approaches to each other on scale-free
networks. We generate networks with scale-free distributions $p(k)\sim k^{-\gamma},k\geq k_{0}$
using the configuration model \citep{newman_random_2002-1}. To ensure
that the networks are connected with high probability, we choose a
minimal degree $k_{0}=3$. We then perform simulations of the SIRS
model for various values of $\lambda$. The epidemic threshold is
determined by finding the maximum of the susceptibility

\begin{equation}
\chi=N\frac{\left\langle \left(\rho^{I}\right)^{2}\right\rangle -\langle\rho^{I}\rangle^{2}}{\langle\rho^{I}\rangle}\,,\label{eq:SIRS_susceptibility}
\end{equation}
where the average is computed in the endemic state, see e.g. \citep{pastor-satorras_epidemic_2015}.

Figure \ref{fig:SIRS_epidemic_threshold} shows a comparison of the
three different predictions to simulations. We first compare the onset
of activity in the endemic state $\rho_{\infty}^{I}$, to TAP, degree-based
mean-field (DBMF) and mean-field predictions in \prettyref{fig:SIRS_epidemic_threshold}
a),b) for different system sizes and varying $\gamma$. We find that
the TAP predictor most closely tracks the onset of the regime with
finite endemic activity. We also compute $\chi$ for the same configurations,
and find that it peaks at the same values of $\lambda$ in \prettyref{fig:SIRS_epidemic_threshold}
c), d). We then determine $\lambda_{c}$ from the peak position of
the susceptibility, and compare it to the different predictors for
varying $N$ in \prettyref{fig:SIRS_epidemic_threshold} (e)-(g).
For small system sizes, the TAP theory yields the most accurate predictions.
As the system size increases, degree-based mean-field theory and the
TAP prediction yield more and more similar predictions. For $\gamma=2.3$,
all three theories eventually agree with the simulation, see \prettyref{fig:SIRS_epidemic_threshold}
c). For $\gamma=2.9$, individual-based and degree-based mean-field
theory do not agree, as we expect. However, both the TAP estimator
and the degree-based mean-field result show a good agreement with
the simulation. This analysis also shows that the TAP prediction and
dynamical message passing are not equivalent: dynamical message passing
overestimates the epidemic threshold in the same parameter regime
\citep{castellano_relevance_2018}. We hence see that although our
theory eliminates self-feedback effects up to second order in $\beta$,
it is not equivalent to DMP.

For $\gamma=3.1$, we also find that the TAP theory produces a good
estimate at small system sizes, see \prettyref{fig:SIRS_epidemic_threshold}
g). For large system sizes, we find that the TAP predictor eventually
converges to the result from degree-based mean-field theory, overestimating
$\lambda_{c}$. We expect that this is due to terms of higher order
in $\beta,$ computing higher order corrections to the TAP result
may improve the prediction.

\paragraph{
\begin{figure}[H]
\centering\protect\includegraphics[width=1\linewidth]{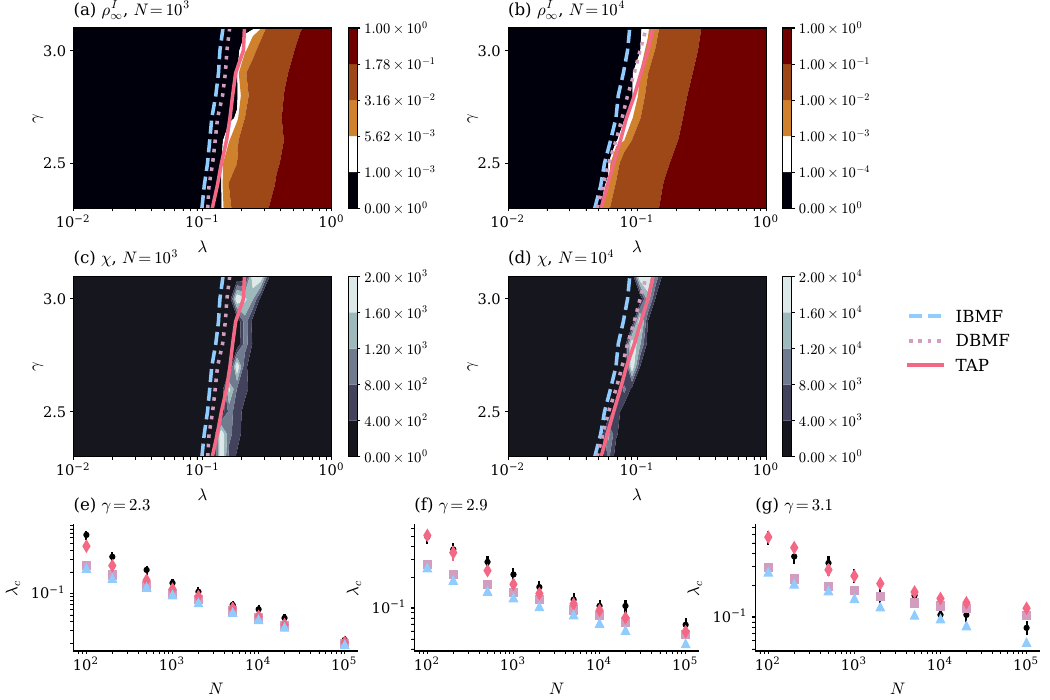}
\protect\caption{\textbf{Epidemic threshold for the SIRS model on scale free networks
with $\eta=0.1$.} \textbf{(a)} Shows the fraction of infected agents
after the dynamics have reached an equilibrium state for varying $\lambda$
and $\gamma$. Dashed curves stem from the mean-field predictions \prettyref{eq:lambda_c_IBMF},
dotted from degree-based mean-field predictions \prettyref{eq:lambda_c_DBMF},
and solid curves indicate the TAP prediction \prettyref{eq:lambda_c_TAP}.
\textbf{(b)} Same as (a), but for larger system size. \textbf{(c)}
Shows susceptibility, \prettyref{eq:SIRS_susceptibility}, for scale
free networks of size $10^{3}$. Curves again indicate the predictions
for $\lambda_{c}$. \textbf{d)} Same as c) but for larger system sizes
$N=10^{4}$. \textbf{e)} Black dots show $\lambda_{c}$ from simulations
over system size for $\gamma=2.3$. Blue triangles, purple squares
and pink diamonds show the predictions from individual-based mean-field
theory, degree-based mean-field theory, and TAP respectively. \textbf{f)}
\textbf{g)} Same as c) but for $\gamma=2.9$ and $\gamma=3.1$, respectively.
All simulation results in panels a)-g) stem from a single realization
of the network connectivity per system size. Theoretical predictions
(mean-field theory, degree-based mean-field theory, and TAP) were
averaged over $10$ different realizations of the connectivity.}
\label{fig:SIRS_epidemic_threshold}
\end{figure}
}

\textbf{}

\bibliographystyle{apsrev4-2}
\bibliography{bib/brain,SIRreferences}

\end{document}